\documentclass[nobalancelastpage,twocolumn,showpacs,nofootinbib]{revtex4-1}
\usepackage{xcolor,amsthm,amsmath,amsfonts,graphicx,bm,amssymb,subfigure}
\usepackage{pifont}
\usepackage{epstopdf}
\usepackage{ulem}
\usepackage{soul}
\usepackage[colorlinks,bookmarks=true,citecolor=blue,linkcolor=red,urlcolor=blue]{hyperref}

\newcommand{\cmark}{\ding{51}}%
\newcommand{\xmark}{\ding{55}}%

\begin{document}
\title{Signatures of topology in quantum quench dynamics and their interrelation}

 \author{Lorenzo Pastori}
 \affiliation{Institute of Theoretical Physics, Technische Universit\"at Dresden, and W\"urzburg-Dresden Cluster of Excellence ct.qmat, 01062 Dresden, Germany}

\author{Simone Barbarino}
\affiliation{Institute of Theoretical Physics, Technische Universit\"at Dresden, and W\"urzburg-Dresden Cluster of Excellence ct.qmat, 01062 Dresden, Germany}
 
 \author{Jan Carl Budich }
 \affiliation{Institute of Theoretical Physics, Technische Universit\"at Dresden, and W\"urzburg-Dresden Cluster of Excellence ct.qmat, 01062 Dresden, Germany}

\begin{abstract}
Motivated by recent experimental progress in the study of quantum systems far from equilibrium, we investigate the relation between several dynamical signatures of topology in the coherent time evolution after a quantum quench. Specifically, we study the conditions for the appearance of entanglement spectrum crossings, dynamical quantum phase transitions, and dynamical Chern numbers. For noninteracting models, we show that in general there is no direct relation between these three quantities. Instead, we relate the presence of level crossings in the entanglement spectrum to localized boundary modes that may not be of topological origin in the conventional sense. We exemplify our findings with explicit simulations of one-dimensional two-banded models. Finally, we investigate how interactions influence the presence of entanglement spectrum crossings and dynamical quantum phase transitions, by means of time-dependent density matrix renormalization group simulations.
\end{abstract}

\maketitle

\section{Introduction}
Founded on the general notion of topological phases of matter~\cite{Hasan_2010,Qi_2011,Wen_17}, physical phenomena reflecting topological properties by now have been predicted and observed in a broad variety of systems.
While in conventional solid-state settings topological states are typically realized at low temperatures, recent advances in quantum simulators, e.g., implemented with ultra-cold atomic gases~\cite{Bloch2008,Goldman_2014,Aidelsburger2018}, provide new opportunities for detecting dynamical signatures of topology in quantum matter far from equilibrium~\cite{Goldman_2016,Eckardt2017,Cooper_2019,Rudner2019}.
There, an enormous tunability enables the implementation of a wide range of topological models \cite{Aidelsburger2011, Sengstock2012,Sengstock2013, Ketterle2013, Aidelsburger2013, jotzu2014, Mancini2015, Stuhl2015} (see \cite{Goldman_2016,Eckardt2017,Cooper_2019} for recent reviews), and the high degree of isolation allows for the realization of coherent quantum many-body dynamics over relatively long timescales.

A common protocol to investigate the interplay between topology and dynamics is to perform a quantum quench, where the system is initialized in a topologically trivial state that can be prepared at low entropy, before some parameters in its Hamiltonian are changed to a topological regime. 
In this scenario, numerous non-equilibrium signatures witnessing the change of topology have been identified~\cite{Bermudez_2009,Foster2013,Rajak_2014,Caio2015,Hu2016,Wilson2016,Sacramento_2016,Wang2017,Barbiero2018,Sun18,Zhang2018,Zache19,Bandyopadhyay19,Hu19,NurUnal2019,McGinley2018,McGinley2019,Heyl2013,Vayna,Budich_2016,Huang_2016,Mendl2019,Sedlmayr,Heyl_2018,Torlai2014,Canovi2014,Gong2018,Yang2018,Flaschner2018,Tarnowski2019,Xu2020}, including dynamical quantum phase transitions~\cite{Heyl2013,Vayna,Budich_2016,Sedlmayr,Huang_2016,Mendl2019,Heyl_2018} (DQPTs), entanglement spectrum crossings~\cite{Torlai2014,Canovi2014,Gong2018} (ESCs), and a dynamical Chern number~\cite{Yang2018} (DCN).
These concepts characterize the postquench time evolution from quite different physical perspectives.
For quench protocols within the same Altland-Zirnbauer (AZ) symmetry class~\cite{Altland97,Schnyder08,Ludwig15}, it is known that DQPTs appear as a consequence of crossing a quantum critical point between a trivial and a topological phase~\cite{Vayna,Budich_2016,Huang_2016,Mendl2019}: 
In this context, several features related to the existence of DQPTs and witnessing the change in the Hamiltonian topology have been experimentally measured, in particular in quench dynamics of two-dimensional Chern insulators \cite{Flaschner2018,Tarnowski2019}.
By contrast, ESCs are a quantum information signature generalizing the presence of protected boundary modes in the entanglement spectrum~\cite{Fidkowski10,Turner_2010,Pollmann10,Berg_11}, thus representing an instantaneous property of the time-evolved state \cite{McGinley2019}. 
Instead, the DCN is a topological invariant defined over a dimensionally extended space-time domain \cite{Yang2018}.

In this work, we present a comprehensive study of the relations between DQPTs, ESCs, and DCN as fingerprints of nonequilibrium topology in quantum quench dynamics, focusing one the fully microscopic study of one-dimensional two-banded systems.
We consider quantum quenches that are not necessarily restricted to a certain AZ symmetry class, and explicitly construct quench protocols exhibiting most of the possible combinations regarding the presence and absence of DQPT, ESC, and DCN (see Table \ref{TABLE}), where the absence of the few unobserved combinations is motivated with a simple geometric picture. 
In this sense, our results imply that there is no one-to-one correspondence between any of pair of those three signatures.

When constraining the quench protocol to a given AZ class, all three signatures individually still constitute a hallmark of some non-trivial topological properties in quench dynamics, albeit an earlier suggested direct correspondence between ESC and DCN~\cite{Gong2018} has been partly refuted ~\cite{McGinley2018,McGinley2019,Lu2019}.
Here, going beyond the notion of symmetry-preserving quenches, we show how these features generally are neither related to topological properties of the pre- and postquench Hamiltonians, nor to an emergent topology of the time-evolved state.
Instead, we observe how these signatures can be dynamically generated even by quenches between topologically trivial Hamiltonians, and how the ESCs can occur as a consequence of accidental boundary modes having no topological origin.
Finally, by means of time-dependent density matrix renormalization group~\cite{white,SchollwoeckReview} (DMRG) simulations, we first investigate the robustness of ESCs and DQPTs against interactions, and finally show how such signatures can appear after a quench of the interaction strength in a correlated version of the Su-Schrieffer-Heeger model \cite{Su_79}.

This paper is structured as follows. In Sec.~\ref{sec:quenchdynamics}, we briefly review how ESCs, DQPTs and DCN can be calculated for two-banded systems out of equilibrium, together with the aspects of out-of-equilibrium topology they relate to. In Sec.~\ref{results} and ~\ref{sec:interaction}  we present our results in the non-interacting and in the interacting regime, respectively. We finally conclude in Sec.~\ref{conclusions}.

\section{Model and Indicators of Topology}
\label{sec:quenchdynamics}
In this section, we introduce the general framework and notations to be used throughout this article. 
We consider a one-dimensional chain of spinless fermions with a number $L$ of unit cells, each one consisting of two orbitals, or sublattice sites, labeled with $A$ and $B$. 
We denote the fermionic operators annihilating a spinless fermion on sublattices $A$ and $B$ with $\hat a_j$ and $\hat b_j$, respectively, where $j=1, \dots, L$. 
We focus on the case where the system is initially --- at time $t=0$ --- prepared in the ground state $|\Psi \rangle$ of a prequench Hamiltonian $\hat H$. 
The Hamiltonian of the system is then suddenly switched to $\hat H'$, and the state of the system will evolve according to $|\Psi(t) \rangle = e^{-i \hat H' t} |\Psi \rangle$. 
At each time $t$ after the quantum quench, the time-evolved state $|\Psi(t) \rangle$ is the ground state of a time-dependent Hamiltonian:
\begin{equation}
\hat H_{\rm P}(t) = e^{-i\hat H't} \, \hat H \, e^{+i\hat H't}  \,,
\label{parent_ham}
\end{equation} 
called {\it parent} Hamiltonian~\cite{Gong2018}, which satisfies the equation of motion $i\partial_t \hat H_{\rm P}(t)=[\hat H', \hat H_{\rm P}(t)]$ with the initial condition $\hat H_{\rm P}(0)=\hat H$. 
From this definition, it is clear that the spectrum of the parent Hamiltonian coincides with that of the prequench Hamiltonian. 
In the following we will mostly focus on non-interacting systems with periodic boundary conditions (PBC): We thus switch to the momentum space description using operators $\hat a_k =\sum_je^{ikj}\,\hat a_j/\sqrt{L}$ and $\hat b_k =\sum_je^{ikj}\,\hat b_j/\sqrt{L}$, with $k=2\pi n/L$ and $n=-L/2,...,L/2-1$ ($L$ is assumed even). 
We denote the parent Bloch Hamiltonian in this basis with $h_{\rm P}(k,t)=\vec{d}_{\rm P}(k,t)\cdot \vec{\sigma}$, with $\vec{\sigma}$ being the vector of the three Pauli matrices acting in the sublattice space, and $\vec{d}_{\rm P}(k,t)$ the Bloch vector. 
Denoting with $\vec{d}(k)$ the Bloch vector for the prequench Hamiltonian $\hat H$, and with $\vec{d'}(k)$ that of the postquench $\hat H'$, the parent Bloch vector $\vec{d}_{\rm P}(k,t)$ can be explicitly calculated as~\cite{Gong2018,Yang2018}: 
\begin{align}
\vec{d}_{\rm P}(k,t)=& \vec{d}_{||}(k)+\cos [2d'(k)t]  \, \vec{d}_{\perp}(k) +\nonumber\\
&+\sin [2d'(k)t] \,  \vec{d}_{\rm o}(k) \,,
\label{DPKT}
\end{align}
where $d(k) \equiv |\vec{d}(k)|$ and $d'(k) \equiv |\vec{d'}(k)|$, and:
\begin{subequations}
\begin{align}
&\vec{d}_{||}(k)=\frac{[\vec{d}(k) \cdot \vec{d'}(k)]}{d'^2(k)} \vec{d'}(k) \label{dparallel} \,, \\
&\vec{d}_{\perp}(k)=\vec{d}(k)-\vec{d}_{||}(k) \,, \\
&\vec{d}_{\rm o}(k)=-\frac{\vec{d}(k) \times \vec{d'}(k)}{d'(k)} \label{d0} \,.
\end{align}
\label{micio}
\end{subequations}
\begin{table}[t]
	\begin{tabular}{c c c| c| c}
		\hline
		\text{ESC} & \text{DQPT} & \text{DCN} & Existence & Protocol \\ \hline\hline
	    yes & yes & yes & \cmark & BDI, D \cite{Gong2018} \\ \hline
		no & yes & yes & \cmark & AIII (dispersive) \cite{Lu2019}, general \\ \hline
		yes & no & yes  & \xmark &  \\ \hline
		no & no & yes & \xmark &  \\ \hline
		yes & yes & no & \cmark & general \\ \hline
		no  & yes & no & \cmark & general  \\ \hline
		yes & no & no & \cmark & general  \\ \hline
		no & no & no & \cmark & general  \\ \hline
	\end{tabular}

	\caption{Summary of our main results in the non-interacting case, showing the relations among ESCs, DQPTs, and DCN, for quenches in one-dimensional two-band models. The \cmark\ or \xmark\ for the rows mark the possibility or impossibility, respectively, of devising quench protocols featuring the corresponding combination of the three signatures in the subsequent unitary time-evolution. The left column shows which protocols could accommodate the corresponding combination: The second combination can e.g.~be found for quenches in AIII class with dispersive post-quench $\hat H'$, and also for general quenches, i.e.~not restricted to a given AZ class.}
	\label{TABLE}
\end{table}
Equipped with the definition of a parent Hamiltonian, we can now review the notions of nonequilibrium topology we will be working with, in the context of quantum quench problems, together with the indicators of the several aspects concerning to it.

\subsection{Topology in quench dynamics}
\label{subsec:quenchtopology}
Here we focus on two {\it inequivalent} definitions of topology in one-dimensional quench problems, and discuss the different topological signatures related to them.
In light of this inequivalence, it is thus not surprising the absence of a general relation between these different indicators.\\
The approach followed by \cite{Gong2018} and \cite{Yang2018} is that of defining the non-equilibrium topology as the $(1+1)$-dimensional topology of the of the parent Hamiltonian, i.e., taking time as an additional dimension.
This is reminiscent of the way topological invariants are defined for Floquet topological insulators \cite{Rudner2019}.
For quench dynamics the periodicity in the time direction is not provided by an external periodic driving, but results from the time-dependence of the parent Hamiltonian \cite{Yang2018}.
In particular, for systems with PBC the parent Hamiltonian has a $k$-dependent time-periodicity $T_k$.
This can be used to define topological invariants as integrals over an extended $k$-$t$ region, whose value does not change when the integration metric is rescaled so as to make $T_k$ independent of $k$: One such invariant is the DCN.
This strategy has been adopted also for two-dimensional band insulators, where this dimensionally extended post-quench topology was quantified in terms of Hopf invariants \cite{Wang2017,Hu19,Tarnowski2019}, and related to the difference of the equilibrium topological numbers of pre- and postquench Hamiltonians.\\
In Refs.~\cite{McGinley2018,McGinley2019} instead, the topology of the time-evolved state is defined as the $1$-dimensional topology of the band-flattened parent Hamiltonian at a fixed $t$, thus concerning only to the properties of the state at a given instant in time.
In this approach, a classification similar to the equilibrium one is then applied to the parent Hamiltonian, based on the symmetries that are dynamically preserved during the time-evolution~\cite{McGinley2018,McGinley2019}.
The presence of level crossings in the entanglement spectrum is what can reveal this aspect of non-equilibrium topology, through bulk-boundary correspondence.\\
For one-dimensional quenches within the same AZ class, looking at the DCN and the entanglement spectrum dynamics thus corresponds to probing the two above definitions of nonequilibrium topology, respectively.
DQPTs instead, occur for quenches crossing a quantum critical point~\cite{Vayna,Budich_2016}.
For a general quench not restricted to a given AZ class, we will show how these three signatures will become unrelated to any topology of the parent, pre- or postquench Hamiltonian.
In the next sections we review the definitions of entanglement spectrum (and its relation to topology), DQPTs and DCN, and how they can be calculated in noninteracting two-band models. \\
Before this, we finally stress that both these definitions of out-of-equilibrium topology are valid only for systems in the thermodynamic limit, or for times not extensive in the size $L$ of the system. 
For a finite system, the parent Hamiltonian at long times would generally be highly nonlocal, reflecting in an increasing number of harmonics in $k$, and we would eventually lack in resolution for any definition of its topology.
This lack of resolution in the long-time limit applies also to the $(1+1)$-dimensional definition of topology, as the momentum-time region chosen for calculating any invariant would scramble due to the generic incommensurability of the periods $T_k$.
Quenches to Hamiltonians having flat bands are in this sense easier to characterize, as they generate a periodic dynamics.

\subsubsection{Entanglement spectrum and its degeneracy}
Given a general many-body state $|\Psi \rangle$ describing a total system, in which a bipartition into a subsystem $S$ and its complement $\bar{S}$ is chosen, the entanglement spectrum (ES) is the set of the eigenvalues $\{\lambda_m \}$ of the reduced density operator $\hat{\rho}_S=\mathrm{Tr}_{\bar{S}}\,|\Psi\rangle\langle\Psi|$.
In the following, we will calculate the entanglement spectrum for the time-evolved state after the quench, choosing as subsystem $S$ the first $\ell=L/2$ real-space unit cells.\\
For noninteracting fermionic systems the entanglement spectrum can be extracted from the knowledge of the single-particle density matrix \cite{Vidal_2003,Peschel_2009}.
We discuss here the relevant aspects of this method, which will help understanding the origin of the degeneracies, or crossings, in the ES.
The single-particle density matrix $C$ has elements $C_{i,j} = \langle\Psi|\,\hat{c}^{\dagger}_i\hat{c}_j|\Psi\rangle$, where $\hat{c}^{\dagger}_i$ denotes a generic fermionic operator creating a state in site/orbital $i$.
Generally, the reduced density operator $\hat{\rho}_S$ can be expressed as $\hat{\rho}_S=e^{-\hat{H}_S}/\text{Tr}\big[e^{-\hat{H}_S}\big]$, where $\hat{H}_S$ is referred to as entanglement Hamiltonian. 
For noninteracting systems, $\hat{H}_S$ can be shown to be a one-body operator, whose single-particle energy levels $\epsilon_n$ are related to the eigenvalues $\xi_n$ of $C$, restricted to subsystem $S$, by the formula $\xi_n=\big(1+e^{\epsilon_n}\big)^{-1}$ \cite{Vidal_2003,Peschel_2009}.
The eigenvalues $\lambda_m$ of the ES are then calculated by specifying the filling of the single-particle entanglement modes $\epsilon_n$. 
Importantly, an eigenvalue $\xi_n=1/2$ for $C$ physically means that the $n$-th particle in the Slater determinant $|\Psi \rangle$ can, with equal probability $1/2$, belong to $S$ or not.
Thus, for every Schmidt eigenstate of $\hat{\rho}_S$ in which the particle is in $S$, there must be another equally probable one where the particle is in $\bar S$, implying all the $\lambda_m$ being degenerate.\\
Half-eigenvalues in $C$ appear when $|\Psi \rangle$ is spanned by single-particle orbitals having equal weight on the left and right of a given bond.
We can thus establish a connection between these states and the presence of zero-energy boundary modes of the parent Hamiltonian for $|\Psi\rangle$.
We first notice the relation $Q=I-2\,C$ between $C$ and the band-flattened parent Hamiltonian $Q$, where $I$ denotes the identity matrix (see \cite{Turner_2010}, and Appendix~\ref{ESpectrum} for details).
In light of the previous observation, an eigenstate of $C$ with half weight on $S$ is typically (exponentially) localized near the entanglement cut, thus resulting in a zero-energy boundary mode for $Q$.\\
The presence of states having equal weight on either sides of a bond --- or equivalently of edge modes of the parent Hamiltonian --- may be guaranteed by the symmetries of the problem, in which case the ESCs are topological (in the sense of symmetry-protected, as we focus on one dimension).
However, such orbitals may also occur accidentally, i.e., in absence of symmetries:
We will see an example of this in section \ref{subsec:III_B_1}.\\
Thus, a topological $|\Psi\rangle$ implies a degenerate ES~\cite{Fidkowski10,Turner_2010,Pollmann10,Berg_11}, but not vice versa, as we will demonstrate in the context of dynamics after a quench.

\subsubsection{Dynamical phase transition}
A DQPT is signaled by a nonanalyticity of a rate function $f(t)$ at a certain instant of time $t$, where $f(t)$ is defined as~\cite{Heyl2013}:
\begin{align}
f(t)=-\lim_{L \rightarrow + \infty } \frac{1}{L} \ln [\mathcal{L}(t)]  \;,
\label{def_DPT_text}
\end{align}
associated to the return probability (Loschmidt echo) $\mathcal{L}(t) = |\langle \Psi| e^{-i\hat H't} |\Psi \rangle |^2$, where $|\Psi \rangle$ denotes the initial state, commonly chosen to be the ground state of the prequench Hamiltonian.
For two-band systems it is possible to show that (see Appendix~\ref{DPT}):
\begin{align}
f(t)= - \int_{-\pi}^{\pi}   \frac{dk}{2\pi} \ln \left[ \cos^2 [d'(k)t] + \gamma(k) \sin^2 [d'(k)t] \right] \,,
\label{DPT_2band}
\end{align}
where $\gamma(k)=\big[ \vec{n}'(k) \cdot \vec{n}(k) \big]^2$, $\vec{n}(k)=\vec{d}(k)/d(k)$ and $\vec{n}'(k)=\vec{d'}(k)/d'(k)$.
From Eq.~\eqref{DPT_2band} we immediately observe that DQPTs occur when $\gamma(k^*)=0$ for some momenta $k^*$. 
Furthermore, we notice that the expression for $\gamma(k)$ is independent of the direction of the quench, and so is the presence of DQPTs.\\
For quenches in the same AZ symmetry class, DQPTs were shown to be related to the change of topological properties of the Hamiltonians before and after the quench~\cite{Vayna,Budich_2016}.
In particular, in~\cite{Budich_2016} a dynamical topological order parameter unambiguously capturing such change was introduced, based on the winding of the Pancharatnam geometric phase in momentum space, which exhibits vortices in correspondence of DQPTs.
Such vortices have been measured in cold atoms experiments, and their evolution characterized in terms of linking numbers and connected to the difference in the topology of pre- and postquench Hamiltonians \cite{Flaschner2018,Tarnowski2019,Xu2020}.
The dynamical topological order parameter has furthermore been measured in the context of quantum walks implemented with photonic systems \cite{Xu2020}.\\
In the following we will also provide example of DQPTs occurring after quenches between topologically trivial Hamiltonians.

\subsubsection{Dynamical Chern number}
The dynamical Chern number is defined for noninteracting models in one dimension, as the Chern number of the parent Hamiltonian in a $(1+1)$-dimensional momentum-time domain. 
The DCN associated to a generic parent Hamiltonian $h_{\rm P}(k,t)= \vec{d}_{\rm P}(k,t) \cdot \vec{\sigma}$ is defined as~\cite{Yang2018} :
\begin{align}
&C^{(m)}_{\rm dyn}= \int_{A_m} \frac{dk}{4\pi}  \int_0^{T_k} dt \,
\vec{n}_{\rm P}(k,t) \cdot \left[\partial_k \vec{n}_{\rm P}(k,t) \times \partial_t \vec{n}_{\rm P}(k,t) \right] \,,
\label{DCNN}
\end{align}
measuring the number of times that $\vec{n}_{\rm P}(k,t)=\vec{d}_{\rm P}(k,t)/|\vec{d}_{\rm P}(k,t)|$, with $\vec{d}_{\rm P}(k,t)$ defined in Eq.~\eqref{DPKT}, would cover the Bloch sphere in a reduced momentum-time manifold.
This reduced manifold is defined by the period of the time evolution for the Bloch states with given momentum $k$, $T_k=\frac{\pi}{d'(k)}$, and the momentum interval $A_m \equiv [k_m, \, k_{m+1}]$, with $m=1, \dots, N$, delimited by the two consecutive momenta $k_m$ and $k_{m+1}$ inside the first Brillouin zone $[-\pi,\pi)$ for which $\vec{n}(k_m)$ is parallel or anti-parallel to $\vec{n}'(k_m)$: 
For each of the $k_m$, $\vec{d}_{\rm P}(k_m,t)$ is constant in time equal to its initial value $\vec{d}(k_m)$ (see Eqs.~\eqref{dparallel}-\eqref{d0}). 
Physically, $\vec{n}'(k_m)$ being antiparallel to $\vec{n}(k_m)$ means that at $k_m$ a band-inversion during the quench has occurred.
Because of the periodicity of $\vec{n}_{\rm P}(k,t)$, one can rescale time to $t=t'/d'(k)$ --- as long as $d'(k)\ne 0$ --- without changing the value of the DCN.
After rescaling, the $N$ momentum-time submanifolds are equivalent to spheres, since the Bloch states at $k_m$ do not evolve apart from a global phase.
Using Eq.~\eqref{DPKT} one can then explicitly evaluate the DCN of Eq.~\eqref{DCNN} to~\cite{Yang2018}:
\begin{align}
&C^{(m)}_{\rm dyn} = \frac12 \left(\cos\theta_{k_m}-\cos\theta_{k_m+1} \right) \,,
\label{dynamical_angle}
\end{align}
with $\theta_{k_m}$ being the angle between $\vec{n}(k_m)$ and $\vec{n}'(k_m)$. 
If no fixed momenta $k_m$ are present, the domain of integration of Eq.~\eqref{DCNN} is equivalent to a torus, in which case the DCN vanishes, as it can be seen from the above equation.
Importantly, in the special case where the quench is performed within the same AZ symmetry class, the DCN can be related to the difference of winding numbers (BDI and AIII class) or of $\mathbb{Z}_2$ topological numbers (D class) of the pre- and postquench Hamiltonians.
This suggests a relation between a finite DCN and the presence of DQPTs, which we will uncover later when presenting our results.
It is worth mentioning that the DCN has been experimentally accessed in recent experiments on quantum walks \cite{Wang2019,Xu2019}.
For general quantum quenches, we will see how a finite DCN can arise quenching between two topologically trivial Hamiltonians, thus demonstrating how this indicator could be dynamically generated also in trivial cases.

\section{Non-Interacting Quench Dynamics}
\label{results}
This section is aimed at discussing the relations between DQPTs, DCNs and ESCs in noninteracting systems. 
First, we discuss the first four cases of Table~\ref{TABLE} where the DCN is different from zero. We show that a finite DCN is a sufficient condition to have a DQPT, while there is no connection between the DCN and the presence of ESCs. 
Then, we address the four remaining cases of Table~\ref{TABLE} characterized by a vanishing DCN and we study a simple quench protocol which allows us to show that there are no connections between ESCs and DQPTs either.
Finally, we show that ESCs are accompanied by the appearance of zero-energy modes and discuss their origin. 
In the following, the vectors $\vec{d}(k)$ and $\vec{d'}(k)$ associated to the prequench and the postquench Hamiltonians respectively fully determine the time evolution of the system. 
As before, we set $\vec{n}(k)=\vec{d}(k)/|\vec{d}(k)|$ and $\vec{n}'(k)=\vec{d'}(k)/|\vec{d'}(k)|$.

\subsection{Non-vanishing dynamical Chern number}

\subsubsection{Simultaneous presence of ESC and DQPT}
\label{YYY}
\begin{figure}
	\begin{center}
  	\includegraphics[width=\columnwidth]{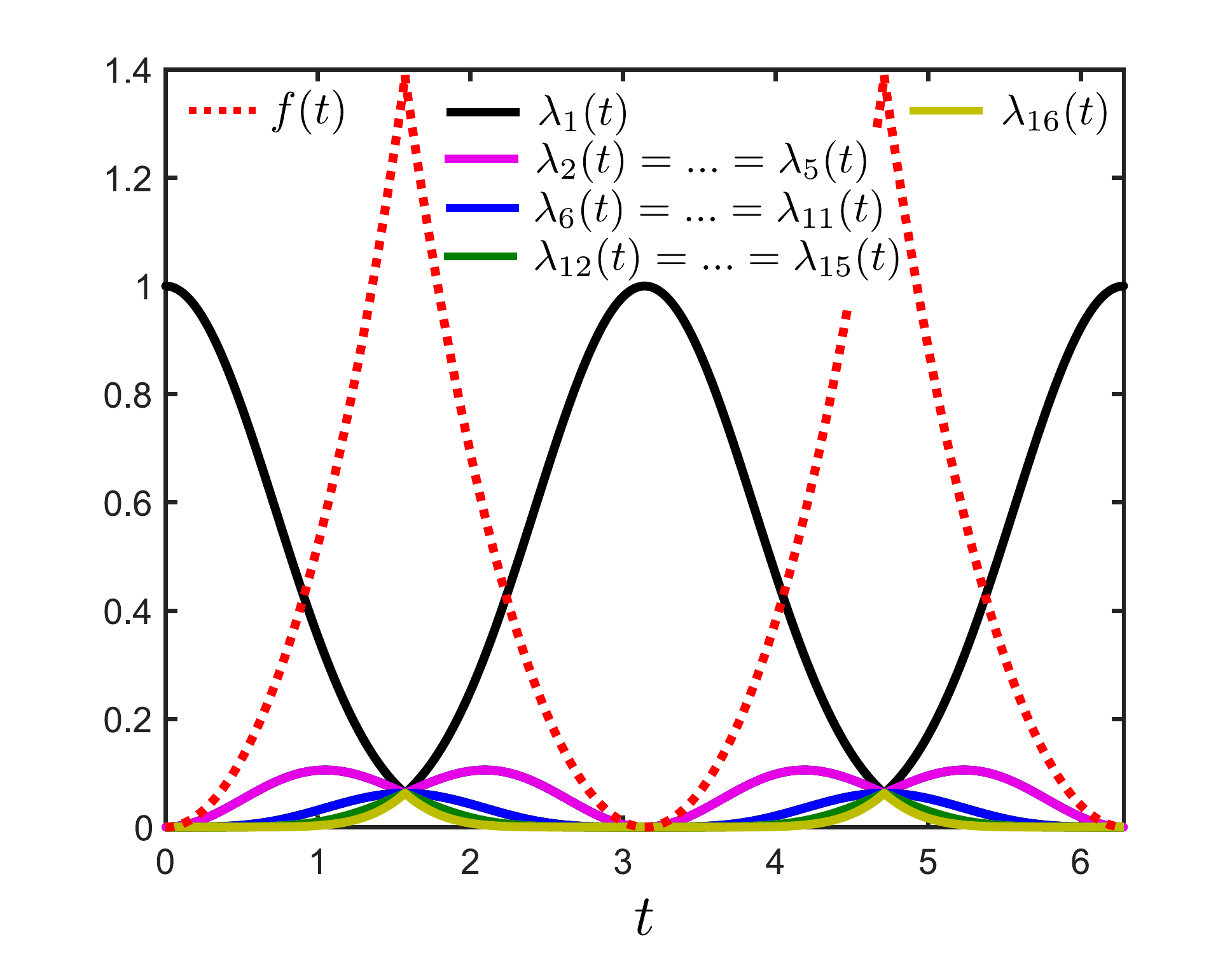}
	\end{center}
	\caption{Time evolution of the entanglement spectrum eigenvalues (solid lines) and of the rate function $f(t)$ (dashed red line) for the quench protocol: $\vec{d}(k)=(J_x,0,0)$ and $\vec{d'}(k)=(J_x\cos k, J_x \sin k , 0)$. Here $L=1000$ sites and $J_x=1$. In this case the ESCs and the DQPTs, signaled by the cusps in $f(t)$, appear at the same instants in time, while the DCN is finite ($\pm 1$).}
	\label{fig1}
\end{figure} 
We consider a system which is prepared in the ground state of a purely classical Hamiltonian and then evolves with the Su-Schrieffer-Heeger (SSH) Hamiltonian~\cite{Su_79}. 
This corresponds to the case studied in Ref.~\cite{Gong2018}. 
We show in the following that the presence of ESCs probes in this case one-dimensional the topology of the parent Hamiltonian $H_P(t)$, and that their degeneracy is related to the number of edge modes in the spectrum of $H_P(t)$.
The Bloch vectors $\vec{d}(k)$ and $\vec{d'}(k)$ corresponding to the prequench and the postquench Hamiltonians are given by:
\begin{subequations}
\begin{align}
&\vec{d}(k)=(J_x,0,0) \,, \label{eq:SSHpre}	\\
&\vec{d'}(k)=(J_x\cos k, J_x \sin k , 0) \label{eq:SSHpost} \,,
\end{align}
\end{subequations}
and are parallel and antiparallel for $k=0$ and $k=\pi$ respectively. 
This identifies two distinct momentum-time regions for the calculation of the DCN, and using Eq.~\eqref{dynamical_angle} we observe that it is quantized to one (minus one) depending on which region we consider. 
The fact that the DCN for the total momentum-time zone sums to zero can be seen as a consequence of the particle-hole symmetry $C=\sigma_z\mathcal{K}$ of the model (with $\mathcal{K}$ being the complex conjugation).
The particle-hole is the only symmetry preserved in the time evolution~\cite{McGinley2018,McGinley2019}, and at all times it relates the parent Bloch vectors $\vec{d}_{\rm P}(k,t)$ and $\vec{d}_{\rm P}(-k,t)$, implying that a positive covering of the Bloch sphere in one half of the momentum-time manifold must appear together with a negative one in the other half.\\
The Bloch vector $\vec{d}_{\rm P}(k,t)$ of the parent Hamiltonian can be determined from Eq.~\eqref{DPKT} (see Appendix~\ref{app_model} for details):
\begin{subequations}
\begin{align}
&{d}_{\rm P}^{(x)}(k,t)= J_x - 2J_x \sin^2 (J_xt) \sin^2 k \,, \\
&{d}_{\rm P}^{(y)}(k,t)=J_x \sin^2 (J_xt) \sin 2k \,, \\
&{d}_{\rm P}^{(z)}(k,t)= -J_x \sin (2J_xt) \sin k \,,
\end{align}
\end{subequations}
which is periodic in time with the same period $\pi/J_x$ for each $k$.
In Fig.~\ref{fig1} (solid lines) it can be seen that at times $t^*=\pi/(2J_x) +m\pi/J_x$ all the eigenvalues in the entanglement spectrum are degenerate.
These ESCs are topological, in the sense that they stem from topologically protected boundary modes in the parent Hamiltonian appearing at times $t^*$, as we discuss in the following.
From the above expression of the parent Hamiltonian it can be easily seen that at times $t^*=\pi/(2J_x) +m\pi/J_x$ it corresponds to a flat-band next-nearest-neighbor SSH model~\cite{Li2014}, i.e., ${d}_{\rm P}(k,t^*)=(J_x\cos 2k,J_x \sin 2k,0)$, with a restored chiral symmetry $S=\sigma_z$, such that $S\,h_{\rm P}(k,t^*)S^\dagger = - h_{\rm P}(k,t^*)$. 
The parent Hamiltonian $\hat{H}_{\rm P}(t^*)$ with open boundary conditions therefore hosts four protected boundary modes at zero energy.
This implies the presence of four zero-energy single-particle entanglement modes, when the half-system bipartition is considered. 
Since the parent Hamiltonian has flat bands, these four entanglement modes are the only ones contributing to the ES (see Appendix~\ref{ESpectrum}).
The number of non-zero eigenvalues in the ES is thus $2^4=16$, and the degeneracy of the four entanglement modes at $t=t^*$ forces all of them to be equal at these times, implying ESCs.\\
The presence of these ESCs is consistent with the out-of-equilibrium classification of Ref.~\cite{McGinley2019}. 
The pre- and postquench Hamiltonians (\ref{eq:SSHpre})-(\ref{eq:SSHpost}) belong to class BDI, possessing time-reversal, particle-hole and chiral symmetry, and can be characterized by a $\mathbb{Z}$ topological invariant (winding number).
Since the only symmetry preserved in the time-evolution is the particle-hole~\cite{McGinley2018,McGinley2019}, the topology of the state out of equilibrium reduces to being classified by a $\mathbb{Z}_2$ invariant (e.g., the Zak phase, as we will define later on).
This $\mathbb{Z}_2$ invariant, denoted with $\nu(t)$ in this section, can be calculated by looking at the real lattice momenta $k=0,\pi$: We can compute it as $(-1)^{\nu(t)}=\text{sign}\big[{d}_{\rm P}^{(x)}(0,t){d}_{\rm P}^{(x)}(\pi,t)\big]$, from which it can be seen that $\nu(t)$ equals $0$, its starting value, at all times. \\
Finally, the existence of DQPTs is inferred by calculating $\gamma(k)=\big[\vec{d}(k) \cdot \vec{d'}(k) \big]^2=J_x^4 \cos^2 k$: For $k=\pi/2$ the system undergoes a DQPT at times $t^*=\pi/(2J_x) +m\pi/J_x$, as can be seen in Fig.~\ref{fig1} (dashed red line).
The fact that in this example DQPTs and ESCs occur at the same times is a consequence of the postquench Hamiltonian having flat bands. 
The presence of band dispersion makes the parent Hamiltonian not periodic in time anymore, and shifts the instants at which DQPTs occur away from the ESCs. This already hints to the fact that DQPTs and ESCs are unrelated.

\subsubsection{Absence of ESCs and presence of DQPTs}
\label{FFF}
\begin{figure}
	\begin{center}
  	\includegraphics[width=\columnwidth]{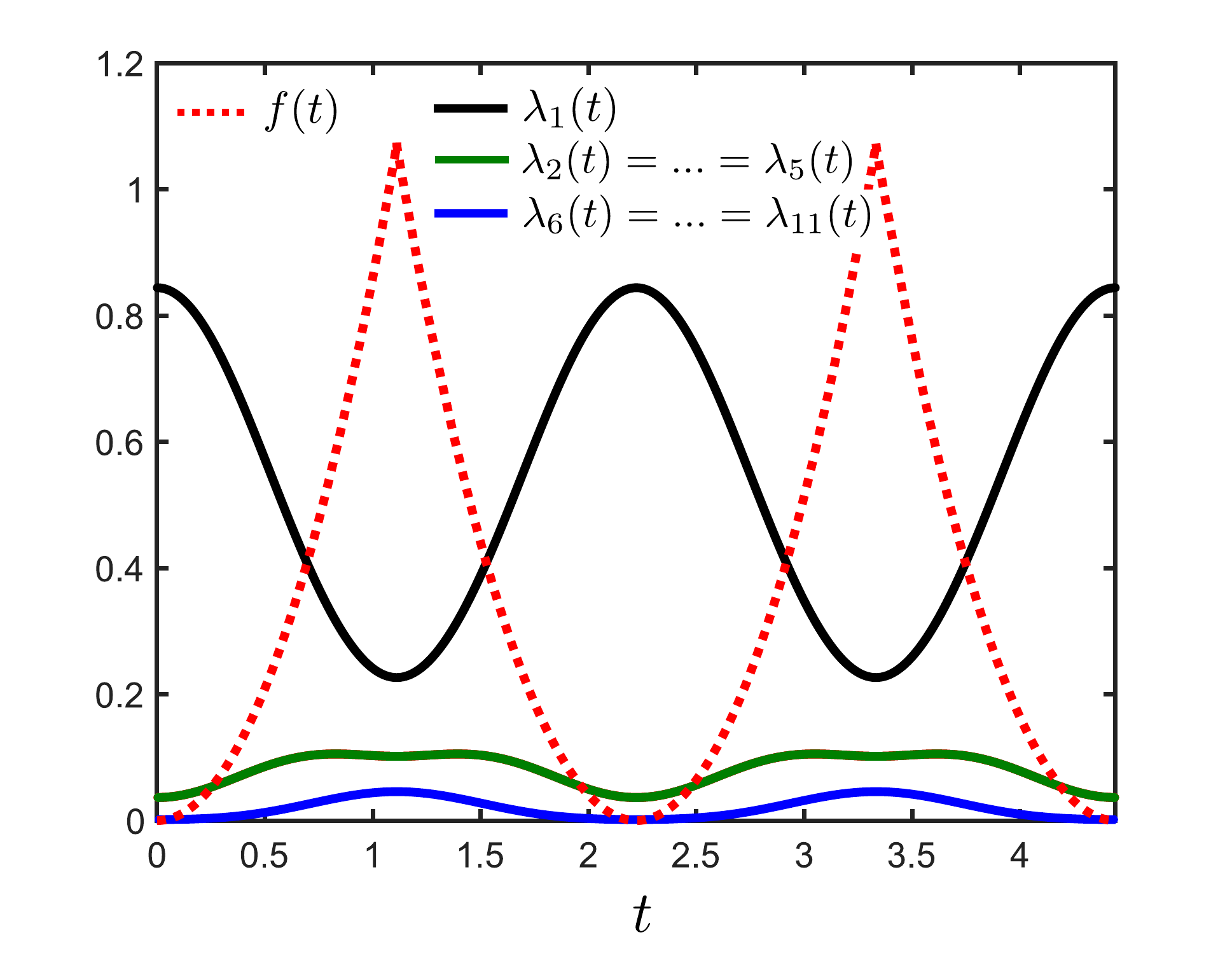}
	\end{center}
	\caption{ Time evolution of the largest entanglement spectrum eigenvalues (solid lines) and of the rate function $f(t)$ (dashed red line) for the quench protocol: $\vec{d}(k)=(J_x,0,J_z \cos k)$ and $\vec{d'}(k)=(J_x \cos k, J_x \sin k , J_z)$. Here $L=1000$, $J_x=1$ and $J_z=1$. In this case, the presence of DQPTs is not accompanied by ESCs, while the DCN is finite ($\pm 1$).}
	\label{fig2}
\end{figure} 
In order to show that a finite DCN does not imply the presence of crossings in the entanglement spectrum, we consider a quantum quench determined by:
\begin{subequations}
\begin{align}
&\vec{d}(k)=(J_x,0,J_z \cos k) \,,	\\
&\vec{d'}(k)=(J_x \cos k, J_x \sin k , J_z) \,.
\end{align}
\end{subequations}
The vectors $\vec{n}(k)$ and $\vec{n}'(k)$ are parallel and antiparallel for $k=0$ and $k=\pi$, respectively, and using Eq.~\eqref{dynamical_angle}, it can be seen that the DCN is quantized to one, again indicating a full winding of the parent Bloch vector $\vec{n}_{\rm P}(k,t)$ around the Bloch sphere in half of the momentum-time zone.
We see that this DCN quantization does not correspond to any topological property of the pre- and postquench Hamiltonian. 
Indeed, the prequench Hamiltonian has a chiral symmetry but it cannot host any topological phase (the winding number is always zero), whereas the postquench Hamiltonian corresponds to a flat-band Rice-Mele model~\cite{RiceMele} with a finite imbalance $J_z$ which prevents the model to have any protecting symmetry. 
As we show in Fig.~\ref{fig2}(a), there are no ESCs, while it is easy to see that DQPTs occurs when $t^*=(2m+1)\pi/\sqrt{J_x^2+J_z^2}$ (see Fig.~\ref{fig2}(b)).\\
It is worth to point out that a similar case (finite DCN and absence of ESCs) can happen also for quenches within the same AZ class.
Considering for example class AIII, in Ref.~\cite{Lu2019} it was discussed how the ECSs can become unstable under band dispersion of the postquench Hamiltonian, despite quenching from a trivial to a topological phase, which implies a quantized DCN~\cite{Yang2018} (and the presence of DQPTs, as we will see below).
This exemplifies the fact that the $(1+1)$-dimensional topology of the parent Hamiltonian, measured by the DCN, in general does not reflect its $1$-dimensional topology, which via bulk-boundary correspondence would become apparent as a degenerate entanglement spectrum.

\subsubsection{Absence of DQPTs}
This case cannot exist. 
In order to have a non-zero quantized DCN, we must have a fixed $k^*$ inside the Brillouin zone such that  $\vec{n}(k)$ and $\vec{n}'(k)$ are parallel, i.e., $\vec{n}(k^*)\cdot \vec{n}'(k^*) = 1$, and a fixed $k^{**}$ such that $\vec{n}(k)$ and $\vec{n}'(k)$ are antiparallel, i.e., $\vec{n}(k^{**})\cdot \vec{n}'(k^{**}) = -1$. 
However, since the function  $\vec{n}(k) \cdot \vec{n}'(k)$ must be continuous there must be a $\overline{k}$ such that 
$\vec{n}(\overline{k}) \cdot \vec{n}'(\overline{k})=\vec{d}(\overline{k}) \cdot \vec{d'}(\overline{k})=0$.  
This last equality implies the existence of a DQPT. For this reason, we observe that a quantized DCN is a sufficient (but not necessary) condition to have a DQPT.

\subsection{Vanishing dynamical Chern number}
\label{subsecDCN0}
\begin{figure}
	\begin{center}
  	\includegraphics[width=\columnwidth]{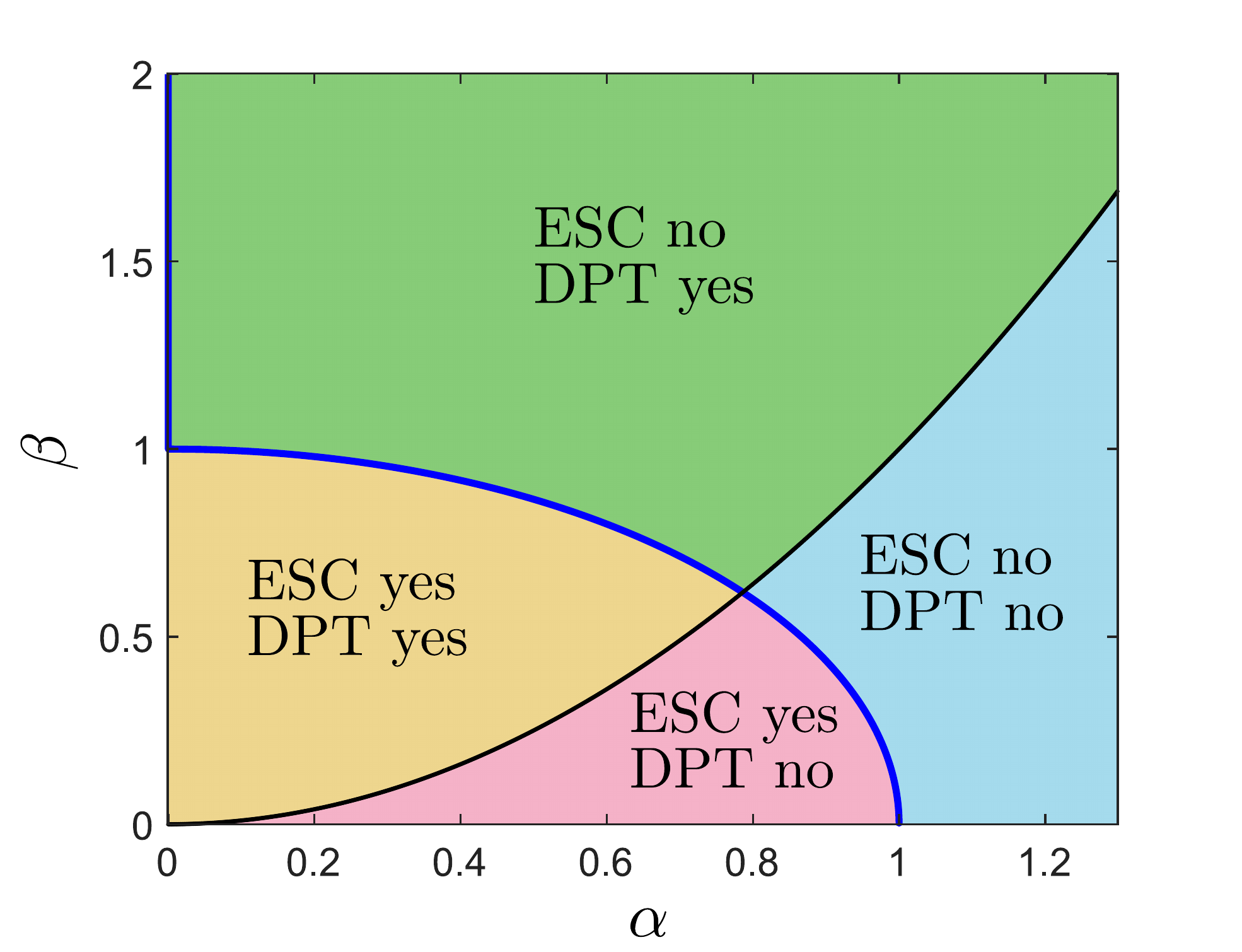}
	\end{center}
	\caption{For the quench protocol $\vec{d}(k)= J(\beta, 0,\alpha)$ to $\vec{d'}(k)= J(\cos k, \sin k, \alpha)$, the parameters $\alpha$ and $\beta$ determine the existence of four regions where DQPTs and/or ESCs appear or do not appear. The existence of ESCs extends to the blue line as well (and thus for any $\alpha=0$), which corresponds to che case of DCN quantized to one.}
	\label{fig3}
\end{figure} 
The last four cases reported in Table \ref{TABLE} can be addressed by studying quantum quenches described by the following Bloch vectors: 
\begin{subequations}
\begin{align}
&\vec{d}(k)=(J_x,0,J_z) \equiv J(\beta, 0,\alpha) \,, \label{eq:noDCN_pre}	\\
&\vec{d'}(k)=(J\cos k, J \sin k, J_z) \equiv J(\cos k, \sin k, \alpha) \label{eq:noDCN_post} \,,
\end{align}
\label{QQQ}
\end{subequations}
where we set $J_x = \beta J$ and $J_z=\alpha J$, with $\alpha>0$ and $\beta>0$ being dimensionless parameters. 
Since the vectors $\vec{n}(k)$ and $\vec{n'}(k)$ can never be antiparallel, the DCN is vanishing for all $\alpha$ and $\beta$.
This can be also checked by explicitly computing the DCN from Eq.~(\ref{DCNN}), integrating over the whole momentum-time zone $[-\pi,\pi)\times[0,T)$ with $T=\frac{\pi}{J\sqrt{1+\alpha^2}}$: The result vanishes for any $\alpha$ and $\beta$, reflecting the fact that the parent Bloch vector $\vec{n}_{\rm P}(k,t)$ does not wrap around the Bloch sphere.
DQPTs occur (do not occur) when $| {\alpha^2}/{\beta} |<1$ ($| {\alpha^2}/{\beta} | >1$). 
As shown in Fig.~\ref{fig3}, DQPTs and ESCs are completely independent. 
Indeed, by varying the parameters $\alpha$ and $\beta$, there exist four regions where any combination of the presence or absence of these features appears.

\subsubsection{Entanglement spectrum crossings and zero-energy modes}
\label{subsec:III_B_1}
\begin{figure}
	\begin{center}
  	\includegraphics[width=0.9\columnwidth]{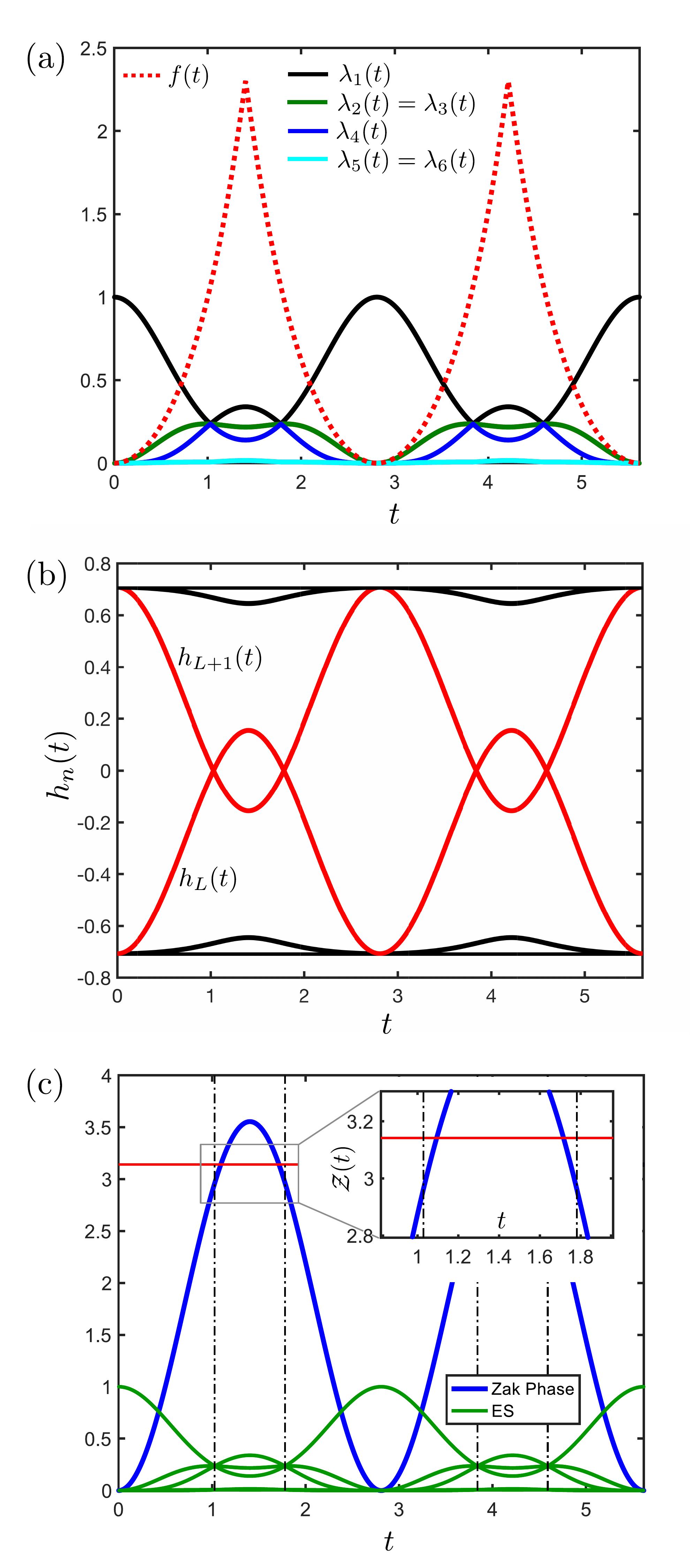}
	\end{center}
	\caption{Dynamics of various signatures for the quench protocol: $\vec{d}(k)= J(\beta, 0,\alpha)$ and $\vec{d'}(k)= J(\cos k, \sin k, \alpha)$ with $\alpha=\beta=0.5$ and $J=1$. 
	(a) Time evolution of the entanglement spectrum $\lambda_m(t)$ (for $m=1,...,6$ --- solid lines), and of the rate function $f(t)$ (dashed red line).
	(b) Time evolution of the eigenvalues $h_n(t)$ of the parent Hamiltonian $\hat{H}_P(t)$ with open boundary conditions. A pair of zero-energy modes appears when the entaglement spectrum is degenerate. 
	(c) Time evolution of the Zak phase $\mathcal{Z}(t)$ of the parent Hamiltonian $h_{\rm P}(k,t)$ (solid blue line). The Zak phase does not equal $\pi$ (marked by the red horizonal line) at the times of the ESCs (marked by vertical dashed black lines --- in green the entanglement spectrum for clarity), implying the absence of inversion as well as other possible protecting symmetries in the AZ sense.}
	\label{fig4}
\end{figure} 
For a proper choice of the parameters $\alpha$ and $\beta$ the time-evolved state after the quench protocol defined by Eqs.~\eqref{eq:noDCN_pre}-\eqref{eq:noDCN_post} exhibits ESCs at certain times. 
In this section, we investigate their origin and relation with zero-energy modes of the parent Hamiltonian.  
Preliminarily, we observe that for $\alpha=0$ and $\beta=1$ we recover the phenomenology studied in Subsection \ref{YYY}, with non-zero DCN. 
In particular, the emerging ESCs are associated to the four zero-energy modes of a generalized SSH model with next-nearest-neighbor hopping terms, which are protected by a chiral symmetry. 
This result can be easily extended to the case $\beta \neq 1$. \\
Interestingly, for $\beta=0$ and $\alpha<1$, the emerging ESCs are associated to the zero-energy modes of a SSH model with nearest-neighbor hopping terms (see Appendix~\ref{app_model}, Fig.~\ref{fig:S1}):
In this case the DCN is zero, but we obtain a nontrivial one-dimensional topology, protected by an emergent chiral symmetry at certain instants in time, despite quenching between two topologically trivial models.\\
We now discuss the ESCs appearing for $\alpha \neq 0$ and $\beta \neq 0$ and we show that such crossings capture the presence of {\it accidental} zero-energy modes, that are not associated to a standard symmetry protected topological phase. 
We proceed as follows. 
We first calculate the vector $\vec{d}_{\rm P}(k,t)$ from Eq.~\eqref{DPKT} for the quench of Eq.~\eqref{QQQ}, and the corresponding parent Hamiltonian $h_{\rm P}(k,t)$ can be written as:
\begin{align}
h_{\rm P}(k,t)= h^{\;(\rm n.)}_{\rm P}(k,t)+h^{\;(\rm n.n.)}_{\rm P}(k,t) \,,
\end{align}
with:
\begin{subequations}
\begin{align}
h^{\;(\rm n. n.)}_{\rm P}(k,t) = \left( \begin{matrix}
0 & \eta(t) e^{-2ik}\\
 \eta^*(t) e^{2ik} &0
\end{matrix} \right) \,,
\end{align}
\begin{align}
h^{\;(\rm n.)}_{\rm P}(k,t) = \left( \begin{matrix}
M(k,t) & \delta(t) + \epsilon(t) e^{-ik}\\
\delta^*(t) + \epsilon^*(t) e^{ik} &-M(k,t)
\end{matrix} \right) \,,
\end{align}
\end{subequations}
with $M(k,t)=m(t)+m_c(t) \cos k + m_s(t) \sin k$. 
The explicit expressions of the functions $\delta(t)$, $\epsilon(t)$, $\eta(t)$, and $M(k,t)$ can be found in Appendix~\ref{app_model}. 
The corresponding real-space Hamiltonian contains on-site and nearest-neighbor hopping terms (from $h^{\;(\rm n.)}_{\rm P}(k,t)$) plus next-nearest-neighbor hoppings (from $h^{\;(\rm n.n.)}_{\rm P}(k,t)$).
By diagonalizing $\hat H_{\rm P}(t)$ with open boundary conditions, we observe that zero-energy modes appear in its single-particle spectrum at the times when the ESCs happen. 
This is clear from Fig.~\ref{fig4} where we show the occurrence of ESCs (a) accompanied by the appearance of (pairs of) zero-energy modes (b). 
A natural question which arises is whether these zero-energy modes are associated to a symmetry protected topological (SPT) phase.
In one dimension, SPT phases with zero-energy modes can appear in the presence of a chiral symmetry $S$ such that $S\,h_{\rm P}(k,t) S^\dagger= - h_{\rm P}(k,t)$ (symmetry classes BDI and AIII) or in the presence of a particle-hole symmetry $C$ such that $C\,h^*_{\rm P}(k,t) C^\dagger= - h_{\rm P}(-k,t)$ (symmetry class D). 
The presence of chiral, particle-hole, and inversion symmetries is ruled out by analyzing the time evolution of the Zak phase of $h_{\rm P}(k,t)$, defined as~\cite{Zak1989,Resta2000}:
\begin{equation}
	\mathcal{Z}(t)=i\int_{-\pi}^{\pi}dk\,\langle u_{\rm P}(k;t)|\partial_k\,u_{\rm P}(k;t)\rangle\,\,,
	\label{eq:Zak}
\end{equation}
with $|u_{\rm P}(k;t)\rangle$ denoting the Bloch state of the lower band of $h_{\rm P}(k,t)$.
As shown in Fig.~\ref{fig4}(c), the Zak phase is not quantized when the zero-energy modes appear.
\begin{figure*} 
	\begin{center}
		\includegraphics[width=2\columnwidth]{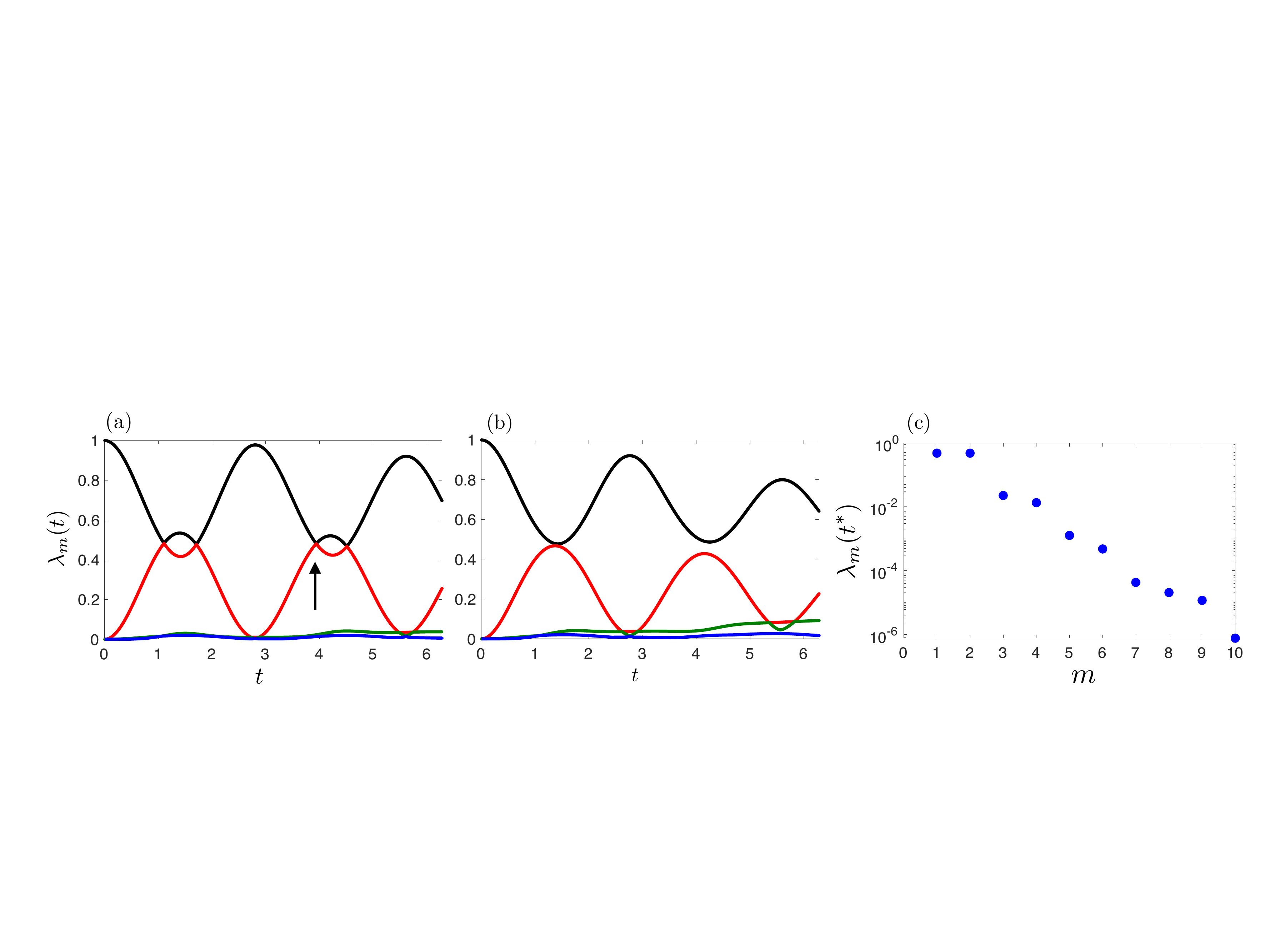}
	\end{center}
	\caption{ Time-evolution of the four largest eigenvalues $\lambda_m(t)$ of the entanglement spectrum for the quench protocol: $\vec{d}(k)= J(\beta, 0,\alpha)$ and $\vec{d'}(k)= J(\cos k, \sin k, \alpha)$ with $\alpha=\beta=0.5$ and different values of the Hubbard interaction term: $U=0.5J$ (a), $U=J$ (b). (c) The entanglement spectrum at a fixed time $t^*$ [see the black arrow in (a)] for which we have crossings between the two largest eigenvalues at $U=0.5J$. Data obtained with DMRG for a chain of $L=96$ unit cells with OBC and $J=1$, using time step $dt=0.0025$ (in units $1/J$). The truncation error is smaller than $10^{-12}$ at all times.}
	\label{fig5}
\end{figure*} 
As long as $\alpha \neq 0$ and $\beta \neq 0$, these emerging zero-energy modes are thus determined by a fine tuning of the parameters in the parent Hamiltonian $h_{\rm P}(k,t)$ and cannot be understood as topological edge states of a SPT phase in one of the AZ symmetry classes.
One can intuitively understand this by first observing that the off-diagonal terms of $h_{\rm P}(k,t)$ describe a generalized SSH model with nearest- and next-nearest-neighbor terms, and a chiral symmetry $S=\sigma_z$. 
Assuming for the moment the diagonal terms to be equal to zero, the generalized SSH model can host four, two or zero exponentially localized zero-energy modes depending on the values of $\eta(t)$, $\epsilon(t)$, and $\delta(t)$. 
Let us now consider a fixed time $t^*$ at which we have ESCs, e.g. $t^*\approx1.028826$ in the case of $\alpha=\beta=0.5$. 
In this case the off-diagonal terms alone, $\eta(t^*)=1/3$, $\epsilon(t^*)=(1+i)/3$, and $\delta(t^*)=-i/6$, would determine the presence of two zero-energy modes protected by the chiral symmetry $\sigma_z$. 
The presence of diagonal terms breaks this chiral symmetry and generally would split these modes away from zero energy, still preserving their exponentially localized nature. 
However at time $t^*$ the interplay of $m(t^*)=-1/6$, $m_c(t^*)=1/3$, and $m_s(t^*)=-1/3$ constitutes a fine-tuning of diagonal elements which restores the twofold degeneracy of the single-particle spectrum at zero energy, even though no chiral symmetry is present. 
These modes have therefore an underlying topological origin, coming from the generalized SSH model in the absence of the diagonal terms, but their degeneracy at zero energy is a result of a particular choice of symmetry-breaking terms.
The presence of these accidental edge modes of the flat-band parent Hamiltonian directly translates to zero-energy levels of the single-particle entanglement Hamiltonian, resulting in a degenerate ES at times $t^*+nT$ and $(T-t^*)+nT$ with $n\in\mathbb{N}$.\\
We have also investigated how the presence of these ESCs is affected by a finite band dispersion in the post-quench Hamiltonian.
The addition of a (small) constant $\sigma_x$ term in Eq.~\eqref{eq:noDCN_post} spoils the periodicity of the dynamics, but does not destroy the full degeneracy of the ES at the crossings.
We show our results with dispersive post-quench bands in the Appendix (see Appendix~\ref{app_model}, Fig.~\ref{fig:S3}).

\section{Interacting Quench Dynamics}
\label{sec:interaction}
In this section, by means of time-dependent DMRG, we study quenches of one-dimensional two-band models in the presence of a Hubbard repulsive interaction $U \sum_j \hat n_{j,a} \hat n_{j,b}$, with $\hat n_{j,a} =\hat a^\dagger_j \hat a_j$ and $\hat n_{j,b} =\hat b^\dagger_j \hat b_j$.
This section is divided into two paragraphs. 
In the first one we study the effects of repulsive interaction on the dynamics resulting from the above quench protocols, focusing in particular on the one defined in Eq.~\eqref{QQQ}.
First, we show that the Hubbard interaction plays a detrimental role to the presence of DQPTs and ESCs: DQPTs disappear for sufficiently strong $U$, while ESCs do not strictly survive even at small interactions.
Then, we study quenches in an interacting SSH model, showing how sudden changes in the interaction strength can generate DQPTs and ESCs.\\
By adding interactions on top of Gaussian states with emergent single-particle topological properties, as done in the next section, one may expect such properties to be spoiled, as the presence of interactions would generally lead to a thermalization of one-body observables.
We however notice that the degeneracy of the entanglement spectrum is not specific to non-interacting problems:
For correlated quantum states, the ESCs result from non-trivial transformation properties of the Schmidt states under the symmetries of the model \cite{Pollmann10,Berg_11}.
A nonequilibrium topological classification of correlated one-dimensional states has been achieved in \cite{McGinley2019b}: Such out-of-equilibrium topology may thus reflect in ESCs after quenches in genuinely interacting models.\\
We finally point out that for interacting models one cannot define a DCN as above, since the state cannot generally be decomposed in terms of single-particle states parametrized on a periodic manifold.
If the state admits some free-fermion description however, as it is the case in our analysis of the interacting SSH model, then the DCN can still be defined.\\
The following numerical simulations have been performed using the ITensor library [\href{http://itensor.org}{http://itensor.org}].

\subsection{DQPTs and ESCs in the presence of interactions}
\begin{figure}
	\begin{center}
		\includegraphics[width=0.88\columnwidth]{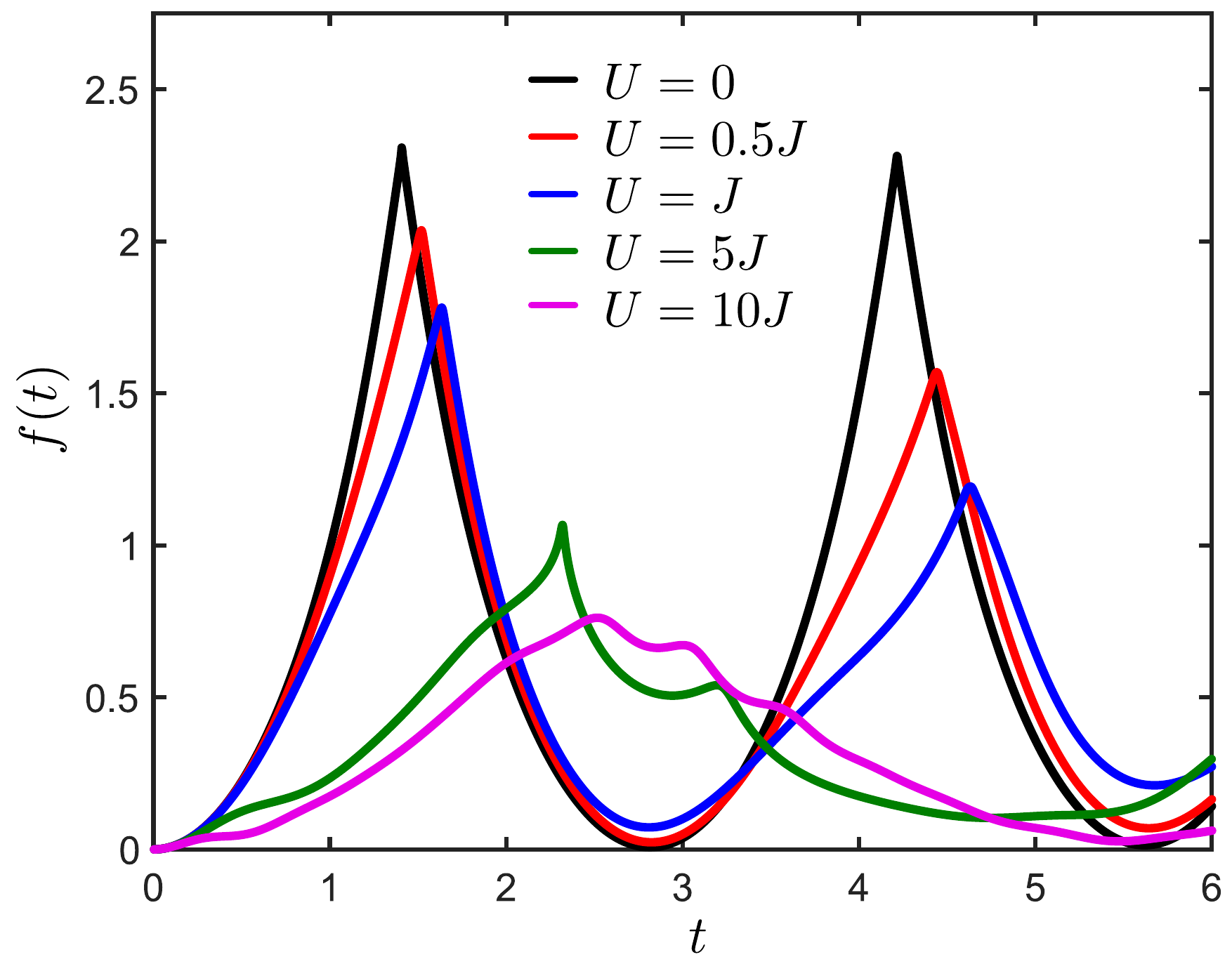}
	\end{center}
	\caption{The rate function $f(t)$ for the quantum quench $\vec{d}(k)= J(\beta, 0,\alpha)$ and $\vec{d'}(k)= J(\cos k, \sin k, \alpha)$ with $\alpha=\beta=0.5$ for different values of the Hubbard interaction term $U$. Data obtained with DMRG for a chain of $L=96$ unit cells with OBC and $J=1$, using time step $dt=0.001$. The truncation error is smaller than $10^{-12}$ at all times. }
	\label{fig6}
\end{figure}

We consider the quench protocol defined in Eq.~\eqref{QQQ} in the regime where $\alpha=\beta=0.5$ such that, in the non interacting case, we have both DQPTs and ESCs. 
We switch on the repulsive Hubbard interaction $U$ in the postquench Hamiltonian, albeit we notice that this is equivalent to the case of having it switched on from the very beginning, it being SU(2) invariant and since we start from a fully polarized state.
In Fig.~\ref{fig5} we investigate the time evolution of the four largest eigenvalues $\lambda_m(t)$ of the ES for different values of the Hubbard interaction $U$ (Panels (a) and (b)). 
Preliminarily, we observe that the time evolution is not periodic anymore, because of the presence of interactions, and many eigenvalues of the entanglement spectrum become non-vanishing during the dynamics.
In the presence of weak repulsive interactions (cf.~Fig.~\ref{fig5}(a)) the ESCs between the two largest eigenvalues are preserved, while they disappear for stronger interactions (cf.~Fig.~\ref{fig5}(b)). 
In Fig.~\ref{fig5}(c), we investigate the degeneracies of the ES at time $t=t^*$ for which we have a crossing between the two largest eigenvalues at $U=0.5J$; cf.~Fig.~\ref{fig5}(a). 
Interestingly, we observe that only the two largest eigenvalues are degenerate, while the lowest ones are not. 
This is in stark contrast with the noninteracting case for which the ES is fully degenerate.
We have verified that this is not a finite size effect by increasing the size of the system up to $L=244$ sites.
We stress that only a full degeneracy of the ES can be associated to topological properties of $|\Psi(t)\rangle$: In this case the time-evolved state becomes trivial in presence of interactions in the postquench Hamiltonian.\\
In Fig.~\ref{fig6} we show the behavior of the rate function $f(t)$ for different values of the interaction term. 
We observe that the divergences of the rate function $f(t)$ persist in the weakly interacting regime, while they disappear for larger interaction strengths, up to the timescales that we have investigated. 
The existence of a critical interaction strength at which DQPTs disappear is not surprising.
Since the initial state is fully polarized, it would not evolve in the limit of post-quench $U\to\infty$, apart from a global phase, the Hubbard interaction being $SU(2)$ invariant.
In this limit, the dynamics is trivial and $f(t)$ vanishes at all times: Thus, there must be a critical interaction strength at which the DQPTs vanish.\\
We finally mention that the phenomenology observed here is qualitatively unaltered if we set $\alpha=0$, namely considering the case of a quench of the form (\ref{eq:SSHpre}) and (\ref{eq:SSHpost}) in the SSH model.
\label{results_interacting}

\subsection{Interacting SSH model}
\label{Ising}
As a different scenario compared to what discussed before, here we consider interaction quenches in an interacting SSH model:
\begin{equation}
\hat H=\sum_{j} \left[ \left(J\hat a^\dagger_{j+1} \hat b_j + J'\hat a^\dagger_{j} \hat b_j +\mathrm{H.c.} \right) + U \hat n_{j,a} \hat n_{j,b} \right] \,,
\label{SSH_interacting}
\end{equation}
which supports at half-filling (the particle number $N$ is equal to the number of sites $L$) a symmetry protected topological phase for $U<U_c$ and a trivial phase for $U>U_c$ with $U_c=4J$ when $J'=0$. The full phase diagram has been studied in Ref.~\cite{Bermudez}. 
In the following, for simplicity, we set $J'=0$ and assume to prepare the system in a trivial state by choosing the prequench interaction $U>U_c$, and quench into the topological phase $U'<U_c$. 
This protocol does not have a counterpart in the noninteracting regime, in that the single-particle terms remain constant.
We study the dynamics numerically with open boundary conditions (OBC) and analytically with periodic boundary conditions (PBC).
We start our analysis assuming PBC and we discuss the existence of DQPTs. To this aim, we introduce the bond operators:
\begin{equation}
\begin{cases}
\hat w_{j,+} = \frac{1}{\sqrt{2}} \left(\hat a_{j+1} + \hat b_{j}\right) \\
\hat w_{j,-} = \frac{1}{\sqrt{2}} \left(\hat a_{j+1} - \hat b_{j}\right) \,,
\end{cases}
\label{porcod}
\end{equation}
and define the $\mathfrak{su}(2)$ algebra operators:  
\begin{subequations}
\begin{align}
&\hat T^{(x)}_{j}=\hat w^\dagger_{j,+}\hat w_{j,-}+ \hat w^\dagger_{j,-}\hat w_{j,+}\,, \\
&\hat T^{(y)}_{j}=-i(\hat w^\dagger_{j,+}\hat w_{j,-}- \hat w^\dagger_{j,-}\hat w_{j,+})\,, \\
&\hat T^{(z)}_{j}=\hat w^\dagger_{j,+}\hat w_{j,+}- \hat w^\dagger_{j,-}\hat w_{j,-}\,,
\end{align}
\label{porcod1}
\end{subequations}
satisfying the usual commutation relations $[\hat T^{(\alpha)}_i, \hat T^{(\beta)}_j] = 2i\epsilon_{\alpha \beta \gamma} \hat T^{(\gamma)}_i \delta_{ij}$. 
Then, we can map the Hamiltonian~\eqref{SSH_interacting} onto an Ising chain in the presence of a transverse magnetic field~\cite{Bermudez}:
\begin{equation}
\hat H=\sum_j \left[ J \hat T^{(z)}_j -\frac{U}{4} \hat T^{(x)}_{j-1} \hat T^{(x)}_{j}\right] \,,
\label{ising}
\end{equation}
and observe that the paramagnetic (anti-ferromagnetic) phase of the Ising chain corresponds to the topological (trivial) phase of the interacting SSH model.
\begin{figure}
	\begin{center}
		\includegraphics[width=0.93\columnwidth]{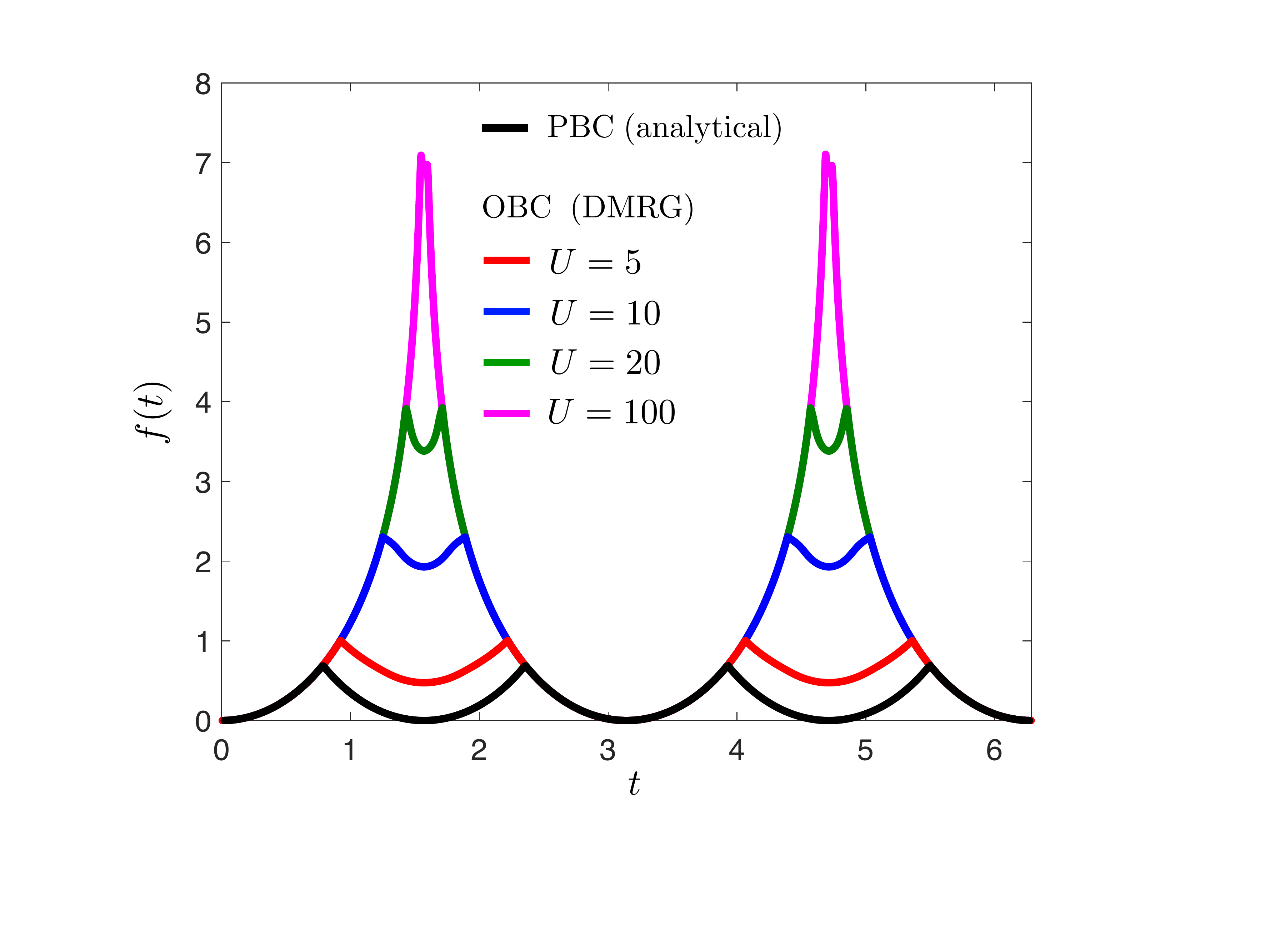}
	\end{center}
	\caption{Time evolution of the rate function $f(t)$ after a quench in the interacting SSH model from finite (large) $U$ to $U'=0$. Black line: $f(t)$ from Eq.~\eqref{DPT_2band}, valid for PBC, for a quench from $U=10J$ to $U'=0$. Red, blue, green and violet lines show $f(t)$ from DMRG simulations with OBC, for a quenches to $U'=0$ starting from $U=5J,\,10J,\,20J$ and $100J$, respectively. In the plot $J=1$, $L=1056$, the time step $dt=0.0025$ (in units $1/J$). The truncation error is smaller than $10^{-12}$ at all times.}
	\label{fig8}
\end{figure} 
Using a standard Jordan-Wigner transformation, the Hamiltonian~\eqref{ising} can be mapped onto a Kitaev chain. 
If we define $\hat c_i = \hat{\mathcal{S}}_{i-1} \hat T^{-}_i $ and $\hat c^\dagger_i = \hat T^{+}_i \hat{\mathcal{S}}_{i-1}  $ with  $\hat T^{\pm}_j=(\hat T^{(x)}_j \pm i \hat T^{(y)}_j)/2$ and $\hat{\mathcal{S}}_{i-1} = \prod_{j=1}^{i-1} e^{i \pi \hat T^+_j \hat T^-_j}$ the usual string operator, we obtain:
\begin{equation}
\hat H=2J \sum_i \hat c^\dagger_i \hat c_i - \frac{U}{4}\sum_i (\hat c^\dagger_i -\hat c_i) (\hat c^\dagger_{i+1} -\hat c_{i+1}) \,.
\label{eq:kitaev}
\end{equation}
Using the Nambu spinors $\hat C^\dagger_k = \left(\hat c^\dagger_k \;\; \hat c_{-k}\right)$ in the momentum space representation, we can rewrite the Hamiltonian~\eqref{eq:kitaev} as $\hat H= \sum_{k>0} \hat C^\dagger_k h(k) \hat C_k$, where  $h(k)= \vec{d}(k) \cdot \vec{\sigma}$, with  $\vec{d}(k)= (0, -U/2 \sin k, 2J-U/2 \cos k)$; similarly the vector $\vec{d'}(k)$ corresponding to the post-quench Hamiltonian is  
$\vec{d'}(k)= (0, -U'/2 \sin k, 2J-U'/2 \cos k)$.  From now on, for simplicity, we set $J=1$. \\
Before addressing the presence of DQPTs, we observe that, despite the presence of interactions, we are able to map the interacting SSH Hamiltonian~\eqref{SSH_interacting} onto a quadratic Hamiltonian. 
Then, using Eq.~\eqref{dynamical_angle}, we can define a DCN for our quench protocol. 
In particular, the DCN is equal to one when the pre- and the postquench Hamiltonian belong to two different phases and vanishes when they belong to the same phase. 
The existence of DQPTs with PBC can be then diagnosed using Eq.~\eqref{DPT_2band} applied to the Kitaev model obtained after the transformations~\eqref{porcod} and~\eqref{porcod1}. 
In the following we assume $U'=0$ and observe that DPTs occur when $t^*=\pi/4+n\pi/2$, see the red line data in Fig.~(\ref{fig8}), as long as $U>U_c$.
\begin{figure}
	\begin{center}
		\includegraphics[width=0.85\columnwidth]{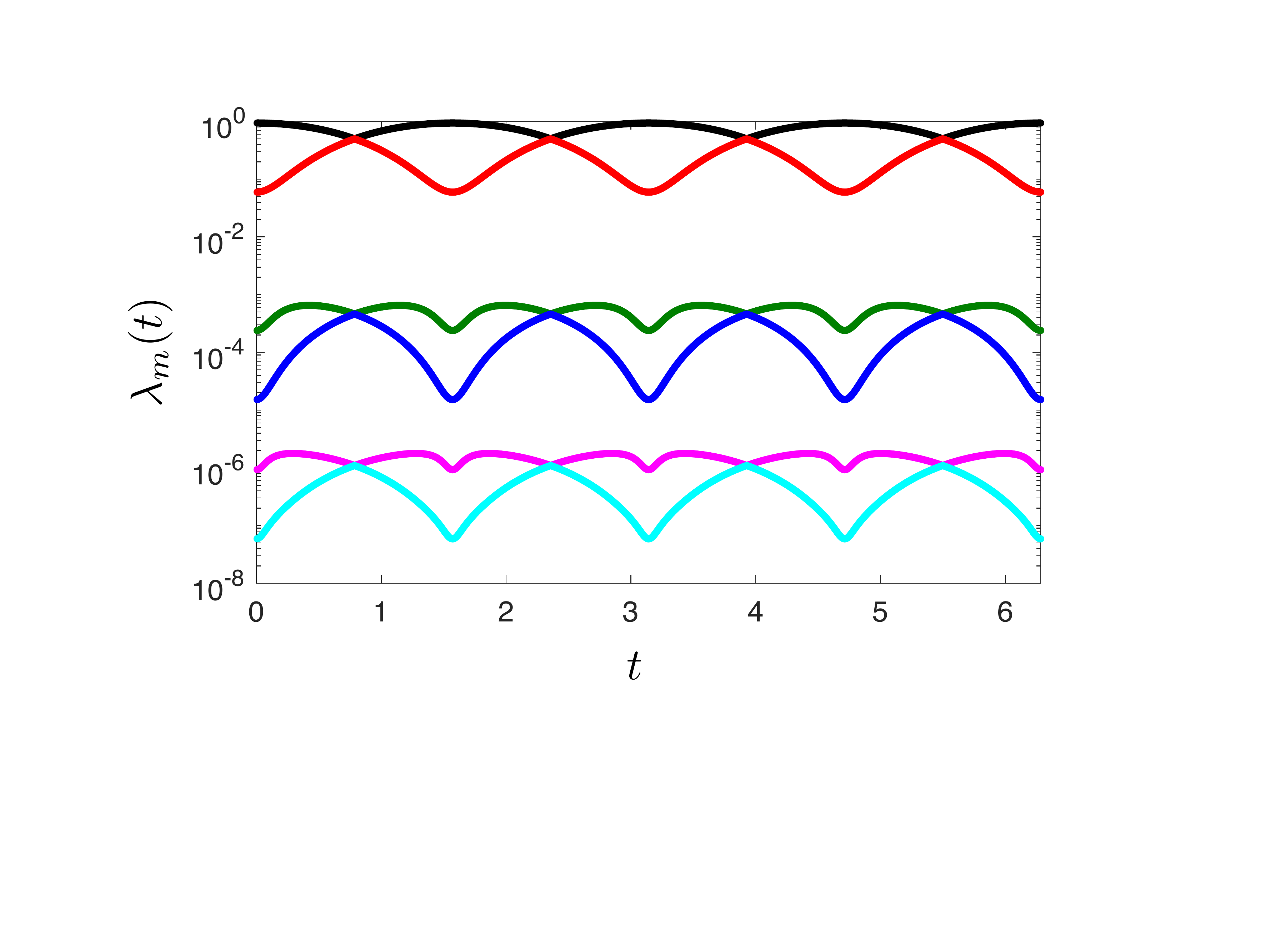}
	\end{center}
	\caption{Time evolution of the six largest eigenvalues $\lambda_m(t)$ of the entanglement spectrum for a quench in the interacting SSH model from $U=5J$ to $U'=0$ with $J=1$. Data obtained with DMRG on a chain of $L=1056$ unit cells with OBC, with time step $dt=0.0025$ (in units $1/J$). The truncation error is smaller than $10^{-12}$ at all times.}
	\label{fig8es}
\end{figure}

\noindent Using time-dependent DMRG with OBC we have studied the behavior of the rate function $f(t)$ defined in Eq.~\eqref{def_DPT_text} by explicitly simulating the quench protocol for the Hamiltonian~\eqref{SSH_interacting} for  $U'=0$ and different values of $U$. 
In Fig.~\ref{fig8} we compare our numerical DMRG data with the analytical prediction obtained using Eq.~\eqref{DPT_2band}. 
We notice that the rate function $f(t)$ obtained from DMRG exhibits DQPTs, but not in correspondence of the times $t^*$ obtained by means of Eq.~\eqref{DPT_2band} which is valid with PBC. 
This discrepancy is a direct consequence of the open boundary conditions and it is not a finite-size effect. 
As explained in detail in Appendix \ref{lorenzo_calc}, the rate function $f(t)$ can be calculated analytically with OBC in the limit where $U' \rightarrow + \infty$, and is expected to exhibit DQPTs at $t^*=\pi/2+n\pi$. 
This is in agreement with our DMRG simulations. 
For smaller values of $U$, DQPTs appear at different times.
Finally, we study the time evolution of the eigenvalues of the entanglement spectrum. 
Since the mapping of the fermionic SSH model Eq.~\eqref{SSH_interacting} in the case $J'=0$ to a transverse-field Ising model~\eqref{Ising} is nonlocal, as it makes use of the bond operators~\eqref{porcod}, the time evolution of the eigenvalues of the entanglement spectrum cannot be understood by exploiting this equivalence between the interacting SSH model and the Ising Hamiltonian. 
Therefore we explicitly simulate the time evolution of the interacting SSH model using DMRG and observe the appearance of entanglement spectrum crossings when quenching from the trivial to the topological phase. 
The results are shown in Fig.~\ref{fig8es}, for a quench from $U=5J$ to $U'=0$.

\section{Concluding discussion}
\label{conclusions}
In summary, we have investigated the relations between entanglement spectrum crossings (ESCs), dynamical quantum phase transitions (DQPTs) and dynamical Chern number (DCN) in one-dimensional two-band models after quenches not necessarily restricted to a given AZ class. 
In the case of noninteracting systems, we devised protocols after which basically any combination between ESCs, DQPTs and quantized DCN can occur --- apart from the case of finite DCN without DQPTs --- see Table \ref{TABLE}. 
While the absence of a one-to-one correspondence between these indicators has been previously discussed in the context of symmetry-preserving quenches, here we showed that for general quenches also their relation to topology is in some cases lost.
In particular, we were able to generate a finite DCN (see Sec.~\ref{FFF}) or topological ESCs (see Sec.~\ref{subsecDCN0} with $\beta=0$ and $\alpha<1$) from topologically trivial Hamiltonians, as well as accidental ESCs related to zero-energy boundary modes not protected by a conventional symmetry.
Going beyond noninteracting systems, we also investigated the robustness of ESCs and DQPTs to the presence of repulsive Hubbard interactions in the considered quench protocols.
While DQPTs are found to persist up to moderate interaction strengths, the full degeneracy of the entanglement spectrum at the crossings does not strictly survive: In this case, interactions are found to destroy the dynamical topology of the state.
Finally, we have considered interaction quenches in a SSH model showing, by means of DMRG simulations complemented by an analytical mapping onto a transverse field Ising chain, how ESCs and DQPTs can arise for such protocols.\\
In this work we have focused on two-band models, which so far have been the most investigated case in experiments.
However, the study of these signatures is not limited to this situation.
DQPTs have been studied in multi-band and disordered systems as well \cite{Huang_2016,Mendl2019}, and a DCN can be defined for quenches in systems with higher number of bands \cite{Gong2018}, in terms of the momentum-time Berry curvature of the parent Hamiltonian. 
In this context, it is interesting to notice that, as in the two-band case studied here, the DCN always vanishes when calculated over the whole Brillouin zone: 
This indicates --- or can rather be seen as a consequence of --- the absence of any topologically protected charge pumping after a single quantum quench experiment.
Regarding ESCs, instances of them after quenches in systems with higher number of bands are also known \cite{Lu2019}.
Generally, topological ESCs stem from a one-dimensional bulk-boundary correspondence in $|\Psi(t)\rangle$, and their existence does not rely on any assumption on the number of bands, neither does the nonequilibrium topological classification in \cite{McGinley2019}.
Thus, via a similar mechanism as in our two-band models, we may expect accidental ESCs to occur in some specific multi-band quenches as well.
A full generalization of our results to systems with higher number of bands remains subject of future work.\\
Other possible future directions include investigating whether there is a more general way of understanding the appearance of the above signatures in quench protocols not constrained to a single AZ class --- which is relevant also beyond the realm of one-dimensional systems --- and studying these aspects of dynamical topology for quenches at finite temperature (see \cite{Heyl2017,Dutta2017} for a discussion in the context of DQPTs).\\
We conclude by pointing out some possible implications of our work in the context of quantum simulation with synthetic materials.
Among the signatures studied here, some --- e.g., DQPTs and the related dynamical vortices in the geometric phase --- have already been experimentally measured and used to detect the topology of post-quench Hamiltonians \cite{Flaschner2018,Tarnowski2019,Xu2020}. 
Furthermore, there are several techniques and proposals that could enable the measurement of entanglement spectra in cold atomic platforms \cite{Cramer2010,Pichler2016,Dalmonte2018,Brydges2019}: The experimental detection of ESCs may thus soon become viable. 
Our results indicate that, for general one-dimensional quantum quenches, one needs to be careful in associating such signatures to some notions of topology. 
As an example one may consider the protocol of section \ref{subsecDCN0}: In absence of lattice imbalance $\alpha$, one would obtain ESCs and DQPTs of topological character, but for a small $\alpha>0$ the presence of these features cannot be traced back to symmetry-protected topological properties in the dynamics anymore.

\acknowledgments
We acknowledge discussions with Jorge Cayao and Max McGinley. L.P.~and J.C.B.~acknowledge financial support from the DFG through SFB 1143 (Project No. 247310070). S.B.~acknowledges the Hallwachs-R\"ontgen Postdoc Program of ct.qmat for financial support. Our numerical calculations were performed on resources at the TU Dresden Center for Information Services and High Performance Computing (ZIH).

\appendix
\section{Non-interacting quench dynamics}
In this Appendix we review the calculation of the entanglement spectrum for non-interacting fermions and show its relation with the topology of the wavefunction. We furthermore show how dynamical quantum phase transitions can be calculated in one-dimensional two-band models, and provide the explicit expression of the parent Hamiltonian for the quench protocol considered in section \ref{subsecDCN0}.

\subsection{Entanglement spectrum and relation to topology} 
\label{ESpectrum}
Here show how to calculate the entanglement spectrum for a noninteracting fermionic system~\cite{Vidal_2003,Peschel_2009}, and how topologically protected boundary modes imply a degeneracy if its levels. 
The physical properties of a non-interacting system can be calculated by using only the single-particle density matrix, which has elements  $C_{i,j}=\langle \Psi |\hat c^\dagger_i \hat c_j |\Psi \rangle$, with $ |\Psi \rangle$ being the state of the system, and $\hat c_j$ being the fermionic operator which annihilate a fermion at site $j$ (including possible spin/orbital degrees of freedom).
In the following we discuss how to calculate the entanglement spectrum from the knowledge of the single-particle density matrix.
The reduced density operator $\hat \rho_S$ for a spatial subsystem $S$ of length $\ell <L$ can be written as: 
\begin{align}
\hat \rho_S = \frac{e^{-\hat H_S}}{Z_S} \hspace{0.2cm} \mathrm{with} \;\; Z_S= \mathrm{Tr}\left[e^{-\hat H_S}\right]  \,,
\end{align}
where $\hat H_S$ is referred to as the entanglement Hamiltonian. 
For non-interacting system all the correlations in $S$ can be calculated from the single-particle density matrix using Wick's theorem: It follows that the entanglement Hamiltonian $\hat H_S$ is quadratic in the fermionic operators, i.e., $\hat H_S = \sum_{i,j=1}^{\ell} \hat c^\dagger_i h^S_{i,j} \hat c_j$.
By introducing the operators $\hat d_m$ which diagonalize $\hat H_S$, i.e. $\hat H_S = \sum_{m=1}^\ell \epsilon_m \hat d^\dagger_m \hat d_m$ with $\hat c_j= \sum_{m=1}^\ell V_{j,m} \hat d_m$, we can express $\hat \rho_S$ as:
\begin{align}
\hat \rho_S = \prod_{m=1}^\ell \frac{e^{-\epsilon_m \hat d^\dagger_m \hat d_m}}{1+e^{-\epsilon_m}} \,,
\end{align}
from which we can see that for $j,k$ being lattice sites in subsystem $S$, we have:
\begin{align}
\mathrm{Tr}_S \left[\hat \rho_S \, \hat c^\dagger_j \hat c_k \right] = \sum_n \frac{V^*_{j,n} V_{k,n}}{1+e^{\epsilon_n}}\equiv(C_S)_{j,k} \,.
\end{align}
This gives us a direct relation between the eigenvalues $\epsilon_m$ of the entanglement Hamiltonian and the eigenvalues of the single-particle density matrix restricted to subsystem $S$, denoted with $C_S$. 
Namely, denoting with $\xi_m$ the eigenvalues of $C_S$, we have:
\begin{align}
\xi_m = \frac{1}{1+e^{\epsilon_m}}\,.
\end{align}
from which we can calculate the entanglement spectrum as:
\begin{align}
\lambda_{\{ s_m \}}= \prod_{m=1}^{\ell} \left[\frac 12 + s_m \left( \xi_m - \frac 12 \right) \right] \hspace{0.2cm} \mathrm{with} \;\;  s_m = \pm 1.
\end{align}
In particular, the highest entanglement eigenvalue would correspond to occupying all the lowest eigenstates of $\hat{H}_S$ with energy $\epsilon_m<0$, i.e. $\xi_m>1/2$, by setting $s_m=1$ for them, and leaving the remaining ones empty with $s_m=-1$. 
The other eigenvalues are simply computed as excitations above this Fermi sea.\\
We show now the relation between the topology of $ |\Psi \rangle$ and the degeneracy of entanglement spectrum \cite{Fidkowski10,Turner_2010}. 
First, we notice that there is a clear correspondence between the single-particle entanglement energies $\epsilon_m$ and the single-particle eigenvalues of the band-flattened parent Hamiltonian for the state $|\Psi\rangle$. 
Since $|\Psi\rangle$ is a Slater determinant of single-particle states $|\varphi_n\rangle$, we may write its band-flattened parent Hamiltonian in first quantization as:
\begin{equation}
	Q = I - 2\sum_{n\;\text{occupied}}|\varphi_n\rangle\langle\varphi_n| \,,
\end{equation}
which can be immediately seen to satisfy:
\begin{equation}
	Q=I-2\,C \,.
\end{equation}
This gives a direct relation between its eigenvalues $q_m$ and the entanglement energies $\epsilon_m$, as:
\begin{equation}
	q_m=1-2\,\xi_m=\frac{e^{\epsilon_m}-1}{e^{\epsilon_m}+1} \,.
\end{equation}
For $q_m=\pm 1$ (i.e., $\xi_m=0,1$) we have $\epsilon_m=\pm\infty$, thus the corresponding entanglement modes are {\it inert}, i.e., they are always empty/occupied and do not contribute to the entanglement spectrum.
If $Q$ has a zero-energy mode with $q_m=0$, the entanglement Hamiltonian also has a single-particle energy mode with energy $\epsilon_m=0$ (corresponding to $\xi_m=1/2$). 
Now let us assume that $|\Psi\rangle$ is the ground state of a topological Hamiltonian $\hat H= \sum_{i,j=1}^L \hat c^\dagger_i H_{i,j} \hat c_j$ that admits protected zero-energy boundary modes with open boundary conditions. 
The entanglement spectrum would be computed by calculating $C$, which is equivalent to computing $Q$ by flattening the bands of $H$.
For local and gapped systems the correlation length is finite, and restricting $C$ (thus $Q$) to subsystem $S$ is not expected to change their topological properties.
That is, the restriction of $Q$ to $S$, denoted with $Q_S$, is topologically equivalent to the band-flattened version of $H$ in $S$, which has topological zero modes, $S$ being a finite subsystem with boundaries.
Since the band-flattening constitutes a smooth deformation, $Q_S$ also has protected edge modes: These result in zero modes in the entanglement energies, and thus degeneracies in the entanglement spectrum.

\subsection{Dynamical quantum phase transitions} 
\label{DPT}
Here we discuss under which conditions a DQPT can appear in two-band systems. 
DQPTs are defined in the main text by Eq.~\eqref{def_DPT_text}, with the Loschmidt echo given by $\mathcal{L}(t) = |\langle \Psi| e^{-i\hat H't} |\Psi \rangle |^2$, where $|\Psi \rangle$ is the initial quantum state. 
Because of the assumed translation-invariance of the model, we can write $\mathcal{L}(t) = \prod_k|\langle u(k)|u_{\mathrm{P}}(k,t)\rangle|^2$, since $|\Psi \rangle$ is a Slater determinant of the Bloch states $|u(k)\rangle$ relative to the lower band of the initial Hamiltonian, and the time-evolved state $e^{-i\hat H't} |\Psi \rangle$ is equivalent to a Slater determinant of the Bloch states $|u_{\mathrm{P}}(k,t)\rangle$ relative to the lower band of the parent Hamiltonian defined in Eq.~\eqref{parent_ham}.
Using the definition~\eqref{def_DPT_text} the rate function reads as: 
\begin{align}
f(t)= - \int_{-\pi}^{+\pi} \frac{dk}{2\pi} \ln |\langle u(k)|u_{\mathrm{P}}(k,t)\rangle|^2 \,,
\end{align}
where we used $\lim_{L \rightarrow +\infty} 1/L \sum_k \approx - \int_{-\pi}^{+\pi} dk/(2\pi)$. 
Defining the projection operators $P(k)=(1-\vec{n}(k)\cdot\vec{\sigma})/2$ and $P_{\rm P}(k,t)=(1-\vec{n}_{\rm P}(k,t)\cdot\vec{\sigma})/2$, with unit Bloch vectors $\vec{n}_{\rm P}(k,t)= \vec{d}_{\rm P}(k,t)/d_{\rm P}(k,t)$ and $\vec{n}(k)= \vec{d}(k)/d(k)$, we have $|\langle u(k)|u_{\mathrm{P}}(k,t)\rangle|^2 = \mathrm{Tr}\left[P(k)P_{\rm P}(k,t)\right]$. Thus:
\begin{align}
|\langle u(k)|u_{\mathrm{P}}(k,t)\rangle|^2 = \frac 12 \left[1+ \vec{n}_{\rm P}(k,t) \cdot \vec{n}(k) \right] \,,
\end{align}
and using Eq.~\eqref{micio} we finally obtain:
\begin{align}
f(t)= - \int_{-\pi}^{+\pi}   \frac{dk}{2\pi} \ln \left[ \cos^2 [d'(k)t] + \gamma(k) \sin^2 [d'(k)t] \right] \,,
\end{align}
with $\gamma(k)=\big[ \vec{n'}(k) \cdot \vec{n}(k) \big]^2$.  
Singularities of $f(t)$ come from those momenta $k^*$ at which $\gamma(k^*)=0$, happening periodically at times $t^*=(2n+1)\pi/d'(k^*)$, with $n\in\mathbb{N}$. \\

\subsection{Parent Hamiltonian for two-band models}
\label{app_model}
Here we show the explicit expression of the parent Hamiltonian for some of the protocols considered in the main text. 
We focus in particular on the case:
\begin{subequations}
\begin{align}
&\vec{d}(k)=J(\beta,0,\alpha) \,,	\\
&\vec{d'}(k)=J(\cos k,  \sin k, \alpha) \,,
\end{align}
\end{subequations}
for which, by direct application of Eqs.~\eqref{dparallel}-\eqref{d0}, we obtain:
\begin{align}
h_{\rm P}(k,t)=\vec{d}_{\rm P}(k,t) \cdot \vec{\sigma}= h^{\;(\rm n.)}_{\rm P}(k,t)+h^{\;(\rm n.n.)}_{\rm P}(k,t) \,.
\end{align}
In the above equation:
\begin{align}
h^{\;(\rm n. n.)}_{\rm P}(k,t) = \left( \begin{matrix}
0 & \eta(t) e^{-2ik}\\
 \eta(t) e^{2ik} &0
\end{matrix} \right) \,,
\end{align}
where:
\begin{equation}
\eta(t)=\frac{\beta J}{1+\alpha^2} \sin^2\big(\sqrt{1+\alpha^2}Jt\big) \,,
\end{equation}
and:
\begin{align}
h^{\;(\rm n.)}_{\rm P}(k,t) = \left( \begin{matrix}
M(k,t) & \delta(t) + \epsilon(t) e^{-ik}\\
\delta^*(t) + \epsilon^*(t) e^{ik} &-M(k,t)
\end{matrix} \right) 
\end{align}
with the off-diagonal terms being:
\begin{subequations}
\begin{align}
\epsilon(t)&=\frac{\alpha J}{1+\alpha^2} \left[\alpha\,\Big(1-\cos\big(2\sqrt{1+\alpha^2} Jt\big)\Big)+\right.\nonumber\\
&\left.+\,i\,\sqrt{1+\alpha^2}\,\sin\big(2\sqrt{1+\alpha^2} Jt\big) \right] \,, \\
\delta(t)&=\frac{\beta J}{2(1+\alpha^2)} \left[1+\big(1+2\alpha^2\big)\cos\big(2\sqrt{1+\alpha^2}Jt\big) +\right.\nonumber\\
&-\left.2i \alpha \sqrt{1+\alpha^2}\,\sin\big(2\sqrt{1+\alpha^2} Jt\big) \right]\,,
\end{align}
and diagonal ones being:
\begin{align}
&M(k,t)=m(t)+m_s(t) \sin k + m_c(t) \cos k \,,\\ 
&m(t)=\frac{\alpha J}{1+\alpha^2}\left[\alpha^2+\cos\big(2\sqrt{1+\alpha^2} Jt\big)\right] \,,\\
&m_c(t)=\frac{2J\alpha \beta}{1+\alpha^2} \sin^2\big(\sqrt{1+\alpha^2} Jt\big) \,,\\
&m_s(t)=-\frac{\beta}{\sqrt{1+\alpha^2}} \sin\big(2 \sqrt{1+\alpha^2}Jt\big) \,.
\end{align}
\end{subequations}

\subsubsection{Case of $\beta=0$ and $\alpha<1$}
\begin{figure}
	\begin{center}
		\includegraphics[width=0.8\columnwidth]{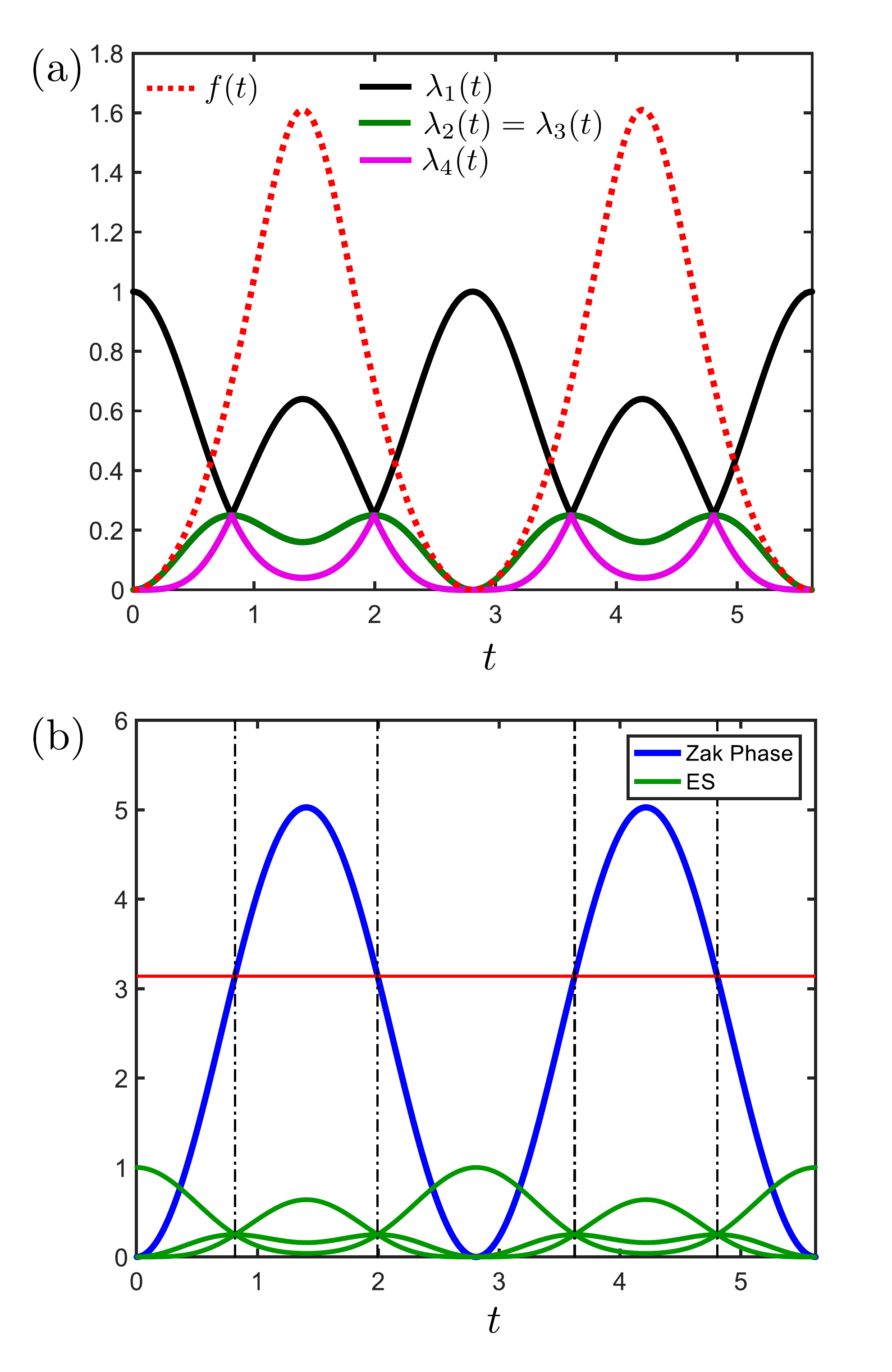}
	\end{center}
	\caption{Dynamics of various signatures for the quench protocol: $\vec{d}(k)= J(0, 0,\alpha)$ and $\vec{d'}(k)= J(\cos k, \sin k, \alpha)$ with $\alpha=0.5$ and $J=1$. 
		(a) Time evolution of the entanglement spectrum $\lambda_m(t)$ (for $m=1,...,4$ --- solid lines), and of the rate function $f(t)$ (dashed red line).
		(b) Time evolution of the Zak phase $\mathcal{Z}(t)$ of the parent Hamiltonian $h_{\rm P}(k,t)$ (solid blue line). The Zak phase equals $\pi$ (marked by the red horizonal line) at the times of the ESCs (marked by vertical dashed black lines --- in green the entanglement spectrum for clarity).}
	\label{fig:S1}
\end{figure}

In the case of $\beta=0$, the coefficients $\eta(t)$, $\delta(t)$, $m_c(t)$ and $m_s(t)$ vanish at all times, and the parent Hamiltonian becomes:
\begin{align}
h_{\rm P}(k,t) = \left( \begin{matrix}
m(t) & \epsilon(t) e^{-ik}\\
\epsilon^*(t) e^{ik} &-m(t)
\end{matrix} \right)  \,.
\end{align}
For $\alpha<1$, there exist time intants at which the diagonal term $m(t)$ vanishes, implying that at this times the parent Hamiltonian has a chiral symmetry $S=\sigma_z$ and a well defined winding number $1$.
At these time instants crossings in the entanglement spectrum occur, as shown in Fig.~\ref{fig:S1}(a), which are associated to the topologically protected edge states of the parent Hamiltonian when OBC are considered.
This can be also seen by looking at the Zak phase, defined in Eq.~\eqref{eq:Zak}, which is equal to $\pi$ when the ESCs happen, as shown in Fig.~\ref{fig:S1}(b).

\subsubsection{Case of $\beta=\alpha=0.5$}
\begin{figure}
	\begin{center}
		\includegraphics[width=0.7\columnwidth]{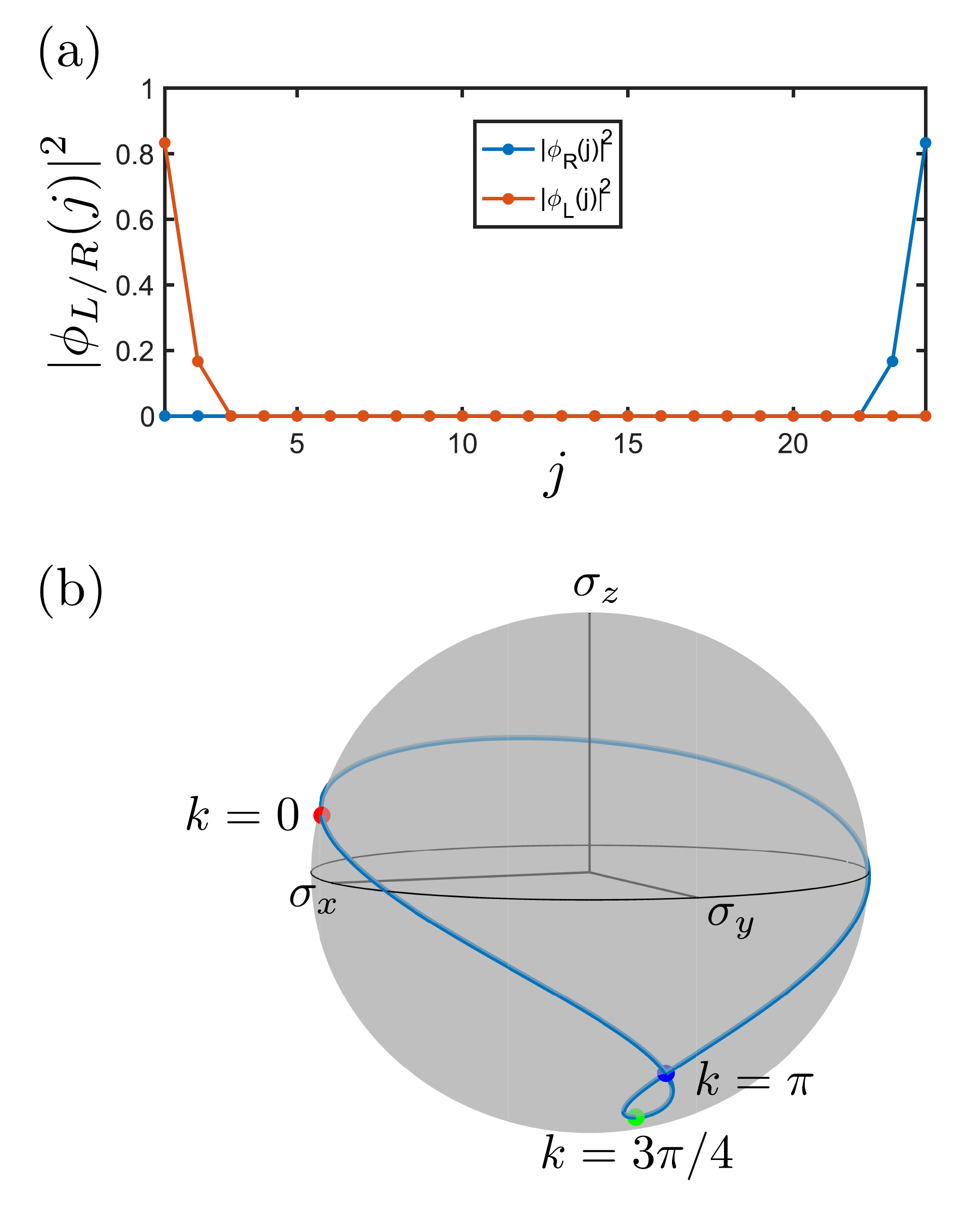}\,\,\,\,
	\end{center}
	\caption{(a) Density profile $|\phi_{L/R}(j)|^2\equiv|\langle j,A|\phi_{L/R}\rangle|^2+|\langle j,B|\phi_{L/R}\rangle|^2$ of the two zero-energy modes $|\phi_{L}\rangle$ and $|\phi_{R}\rangle$ of the parent Hamiltonian (\ref{eqS:parentHamESC}) with OBC. (b) Evolution of $\vec{n}_{\rm P}(k,t_1^*)$ as a function of $k$ in the Brillouin zone.}
	\label{fig:S2}
\end{figure}

\begin{figure}
	\begin{center}
		\includegraphics[width=0.9\columnwidth]{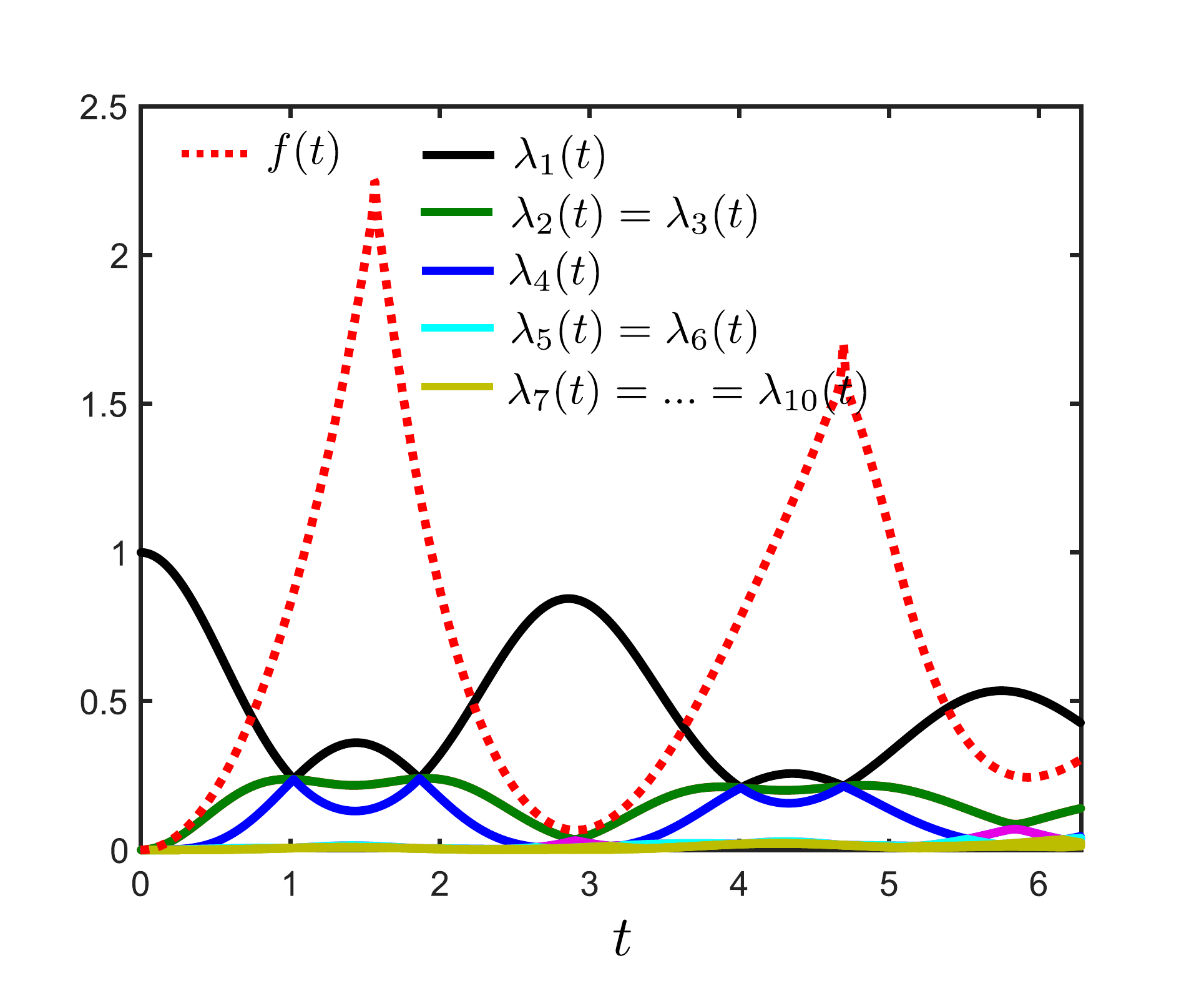}\,\,\,\,
	\end{center}
	\caption{Dynamics of various signatures for the quench protocol: $\vec{d}(k)= J(\beta, 0,\alpha)$ and $\vec{d'}(k)= J(\delta+\cos k, \sin k, \alpha)$ with $\beta=\alpha=0.5$, $\delta=0.2$, and $J=1$.}
	\label{fig:S3}
\end{figure}

In the case of $J=1$ and $\alpha=\beta=0.5$, one can see that, at the time $t_1^*\approx1.028826$ when the ESCs happen for the first time, the Bloch vector for the parent Hamiltonian takes the form: 
\begin{equation}
	\vec{d}_{\rm P}(k,t_1^*)=\gamma\begin{pmatrix}
	2(\cos k + \sin k) + 2 \cos 2k \\
	1 - 2(\cos k - \sin k) + 2 \sin 2k \\
	2(\cos k - \sin k) - 1
	\end{pmatrix} \,,
	\label{eqS:parentHamESC}
\end{equation}
with $\gamma=1/6$, while at the second time $t_2^*$ we have $\vec{d}_{\rm P}(k,t_2^*)=\vec{d}_{\rm P}(-k,t_1^*)$. 
The spectrum of these models with open boundary conditions has a pair of exponentially localized edge modes at zero energy, whose density profile is shown in Fig.~\ref{fig:S2}(a), although no particle-hole, chiral or inversion symmetry is present, which can be seen from the fact that the Zak phase is different from $\pi$ (see Fig.~\ref{fig4}(c)), and also by looking at the evolution of $\vec{n}_{\rm P}(k,t_1^*)=\vec{d}_{\rm P}(k,t_1^*)/d_{\rm P}(k,t_1^*)$ as a function of $k$ in the Brillouin zone in Fig.~\ref{fig:S2}(b). 
In Fig.~\ref{fig:S3} we show the presence of ESCs and DQPTs in the case of $J=1$ and $\alpha=\beta=0.5$ with the addition of small band dispersion, in the form of a constant $\sigma_x$ term in the postquench Hamiltonian.

\section{Interacting quench dynamics}
\label{lorenzo_calc}
In this Appendix, we report details regarding the calculation of dynamical phase transitions in the interacting models considered.

\subsection{DQPT in interacting SSH model} 
Here we consider the quench protocol for the interacting SSH model presented in the Subsection~\ref{Ising} and we discuss how the choice of the boundary conditions affects the times $t^*$ in correspondence of which DQPTs appear.
We recall that the pre- and postquench Hamiltonians are:
\begin{equation}
\hat H=\sum_{j=1}^{L_{\rm BC}} \left(J\hat a^\dagger_{j+1} \hat b_j +\mathrm{H.c.} \right) + U\sum_{j=1}^{L} \hat n_{j,a} \hat n_{j,b} \,,
\end{equation}
and:
\begin{equation}
\hat H'=\sum_{j=1}^{L_{\rm BC}} \left(J\hat a^\dagger_{j+1} \hat b_j +\mathrm{H.c.}  \right) \,,
\end{equation} 
with $L_{\rm BC}=L$ ($L_{\rm BC}=L-1$) for periodic (open) boundary conditions. 
In particular, we discuss the reason why, with open boundary conditions and in the regime $U\gg J$, DQPTs appear at $t^*=\pi/2 + n \pi/2$ (in units of $J$) rather than at $t^*=\pi/4 + n \pi/2$ as expected with periodic boundary conditions.
In the regime $U\gg J$, using standard perturbation theory, it is easy to prove that the the ground state of the pre-quench Hamiltonian is doubly degenerate, and the two lowest energy configurations are given by $|\Psi_A\rangle= \prod_{j=1}^L \hat a_j^{\dagger} |0\rangle$ and $|\Psi_B\rangle= \prod_{j=1}^L \hat b_j^{\dagger} |0\rangle$. 
Because of the inversion symmetry of the SSH Hamiltonian $\hat{I}=\sigma_x\otimes\hat{R}$ (with $\hat{R}$ being the spatial reflection and $\sigma_x$ acting on the sublattice indices), the two degenerate ground states correspond to the symmetric and the antisymmetric linear combinations $|\Psi_\pm \rangle= (|\Psi_A\rangle \pm |\Psi_B\rangle)/\sqrt{2}$. 
Introducing the bond operators: 
\begin{equation}
\begin{cases}
\hat w_{j,+} = \frac{1}{\sqrt{2}} \left(\hat a_{j+1} + \hat b_{j}\right)\\
\hat w_{j,-} = \frac{1}{\sqrt{2}} \left(\hat a_{j+1} - \hat b_{j}\right) \,,
\end{cases}
\end{equation} 
the postquench Hamiltonian takes the form: 
\begin{equation}
\hat H'= J \sum_{j=1}^{L-1} \left(\hat w^\dagger_{j,+} \hat w_{j,+}  - \hat w^\dagger_{j,-} \hat w_{j,-}  \right) \,,
\end{equation} 
consisting of on-site terms only. 
For convenience we also define the single-particle states $|j,A \rangle = \hat a^	\dagger_j|0\rangle$, $|j,B \rangle = \hat b^\dagger_j|0\rangle$ and $|j,+ \rangle = \hat w^\dagger_{j,+}|0\rangle$ and $|j,- \rangle = \hat w^\dagger_{j,-}|0\rangle$. 
We observe that with open boundary conditions, $j=1,..., L-1$ and the states $|1,A \rangle$ and $|L,B \rangle$ are completely decoupled. 
Since $|\Psi_A\rangle$ and $|\Psi_B\rangle$ are Slater determinants, under the action of the noninteracting postquench Hamiltonian $\hat H'$, their single-particle states evolve in time as:
\begin{subequations}
\begin{align}
e^{-i \hat H' t } |j,A \rangle = e^{-i t }  |j-1,+ \rangle -  e^{i t } |j-1,- \rangle \label{QQ} \,,
\end{align}
for $j=1,\cdots, L-1$ and:
\begin{align}
e^{-i \hat H' t } |j,B \rangle = e^{-i t }  |j,+ \rangle + e^{i t } |j,- \rangle \label{QQQQ} \,
\end{align}
\end{subequations}
for $j=2,\cdots, L$. 
More explicitly we observe that, with open boundary conditions, the states $|1,A \rangle$ and $|L,B \rangle$ do not evolve. 
We now consider the time $t^*=\pi/2$  and, in order to diagnose the appearance of a DQPT, we calculate the overlap:
\begin{align}
\langle \Psi_\pm|\Psi_\pm(t)\rangle&= \big[\langle \Psi_A|\Psi_A(t)\rangle+\langle \Psi_B|\Psi_B(t)\rangle \pm \nonumber\\
&+ (\langle \Psi_A|\Psi_B(t)\rangle + \langle \Psi_B|\Psi_A(t)\rangle)\big]/2 \,.
\end{align} 
Using Eqs.~\eqref{QQ}-\eqref{QQQQ}, it is easy to prove that:
\begin{subequations}
\begin{align}
e^{-i \hat H' \frac{\pi}{2}} |j,A\rangle = -i |j-1,B \rangle \,, \\
e^{-i \hat H' \frac{\pi}{2}} |j,B\rangle = -i |j+1,A \rangle \,.
\end{align}
\end{subequations}
Using the fact that for two Slater determinants $|\Psi\rangle$ and $|\Phi\rangle$ made of $L$ single-particle states $|\psi_j\rangle$ and $|\phi_j\rangle$ with $j=1,...,L$, respectively, their overlap is given by the determinant of the single-particle overlap matrix, i.e. $\langle\Phi|\Psi\rangle=\det\big[\{\langle\phi_i|\psi_j\rangle\}_{i,j}\big]$, we observe that $\langle \Psi_A|\Psi_A(\pi/2)\rangle =\langle \Psi_B|\Psi_B(\pi/2)\rangle=0$ with both OBC and PBC.
We thus concentrate on $\langle \Psi_B|\Psi_A(\pi/2)\rangle$ (similar arguments hold for $\langle \Psi_A|\Psi_B(\pi/2)\rangle$). 
In the case of OBC since $|1,A\rangle$ remains unchanged, while all other $|j,A\rangle$ states in $|\Psi_A\rangle$ shift to $B$ states, the overlap matrix has a column of zeros and therefore its determinant vanishes. 
Thus, at $t^*=\pi/2$ we have $\langle \Psi_\pm|\Psi_\pm(\pi/2)\rangle=0$, which implies a DQPT.
On the contrary, with PBC, it is possible to show that $\langle \Psi_\pm |\Psi_\pm(\pi/2)\rangle = \pm (-1)^{L/2}$, which implies that the rate function vanishes.

\subsection{Calculation of rate function with DMRG} 
Here we discuss how to practically calculate the rate function $f(t)$ using matrix product states (MPS) techniques when the system size $L$ becomes large. 
When the initial state $|\Psi\rangle$ and its time evolved $|\Psi(t)\rangle$ are expressed as MPS their overlap can be straightforwardly calculated.
However, at times $t^*$ at which we have a DPT we have that the rate function $f(t)$ defined in Eq.~(\ref{def_DPT_text}) becomes of order one, meaning that $|\langle\Psi|\Psi(t^*)\rangle|^2\sim O(e^{-L})$. 
For large system sizes the value of $|\langle\Psi|\Psi(t^*)\rangle|^2$ is thus so small that it cannot be represented on a computer, thus yielding incorrect results. 
To overcome this problem, we iteratively rescale the overlap during its calculation in the MPS representation, to keep it of order one, and resum the logarithms of the rescaling factors at each step in order to access the rate function at the end of the calculation. 
Specifically, we calculate $\tilde{O}(t)=A(t)\langle\Psi|\Psi(t)\rangle$ with $A(t)=\prod_{j=1}^{L}\alpha_j(t)$, where the $\alpha_j$ are real rescaling parameters chosen such that $|\tilde{O}(t)|\sim O(1)$, and in terms of $\tilde{O}(t)$ the rate function becomes:
\begin{equation}
	f(t) = -\frac{1}{L}\log|\tilde{O}(t)|^2+\frac{2}{L}\sum_{j=1}^{L}\log\alpha_j(t)\,,
\end{equation}
where the first term is now negligible if $|\tilde{O}(t)|\simeq 1$. 
We perform the following choice for the $\alpha_j(t)$. 
For MPS expressed as:
\begin{equation}
	|\Psi\rangle=\sum_{\sigma_1,...,\sigma_L}M^{[1]\sigma_1}M^{[2]\sigma_2}...M^{[L]\sigma_L}|\sigma_1,\sigma_2,...,\sigma_L\rangle\,,
\end{equation} 
with $\sigma_j$ denoting the occupation number at site $j$, and $M^{[j]\sigma_j}$ being matrices with elements $M^{[j]\sigma_j}_{a_{j-1},a_j}$ (with $M^{[1]\sigma_1}$ and $M^{[L]\sigma_L}$ being row and column vectors respectively), and $|\Psi(t)\rangle$ having the same expression with different matrices $\tilde{M}$, the overlap is conveniently calculated by introducing matrices $C^{[j]}$ whose elements are given by:
\begin{equation}
	C^{[j]}_{a_j,a_j'}=\sum_{\sigma_j}\big(M^{[j]\sigma_j}\big)^{\dagger}_{a_j,a_{j-1}} C^{[j-1]}_{a_{j-1},a_{j-1}'} \tilde{M}^{[j]\sigma_j}_{a_{j-1}',a_j'}\,,
\end{equation}
with $C^{[0]}=1$ being the identity~\cite{SchollwoeckReview}. The overlap is then $\langle\Psi|\Psi(t)\rangle=C^{[L]}$. 
We compute the $\alpha_j$ by rescaling the $C^{[j]}$ at each step of the computation of the overlap $C^{[L]}$, that is for each $j$, $C^{[j]}$ is replaced by $\alpha_jC^{[j]}$ where:
\begin{equation}
	\alpha_j=\text{tr}\big[(C^{[j]})^{\dagger}C^{[j]}\big]^{-\frac{1}{2}}\,,
\end{equation}
which keep $|\tilde{O}(t)|$ of order one, therefore avoiding problems related to exponentially small numbers.

\end{document}